\documentclass[traditabstract]{aa} 
\usepackage{graphicx}
\usepackage{txfonts}
\usepackage{wasysym}
\usepackage{array}
\usepackage{natbib}
\bibpunct{(}{)}{;}{a}{}{,} 
\usepackage{ulem}
\usepackage{aas}
\begin{document}
   \title{TASTE. III. A homogeneous study of \\ transit time variations in 
          WASP-3b\thanks{This article is based on observations made with the IAC80 telescope 
          operated on the island of Tenerife by the Instituto de Astrof\'{\i}sica de Canarias (IAC) 
          in the Spanish Observatorio del Teide.}}
   \author{V. Nascimbeni\inst{1,2}\thanks{email address: \texttt{valerio.nascimbeni@unipd.it}}
          \and A. Cunial\inst{1}    \and S. Murabito\inst{1,3} \and P. V. Sada\inst{4}
          \and A. Aparicio\inst{3,5}  \and G. Piotto\inst{1,2}  \and L. R. Bedin\inst{2} 
          \and \\ A. P. Milone\inst{3,5} \and A. Rosenberg\inst{3,5} 
          \and L. Borsato\inst{1,2} \and M. Damasso\inst{1,6}\thanks{INAF associated} \and V. Granata\inst{1} 
          \and L. Malavolta\inst{1}
          }
   \institute{Dipartimento di Astronomia, Universit\`a degli Studi di Padova,
              Vicolo dell'Osservatorio 3, 35122 Padova, Italy
         \and
             INAF -- Osservatorio Astronomico di Padova, vicolo dell'Osservatorio 5, 35122 Padova, Italy
         \and Departamento de Astrof\'{\i}sica, Universidad de La Laguna (ULL), E-38206 La
             Laguna, Tenerife, Canary Islands, Spain
         \and
             Universidad de Monterrey, Departamento de F\'isica y Matem\'aticas, Av. I. Morones
             Prieto 4500 Pte., San Pedro Garza Garc\'{\i}a, Nuevo Le\'on, 66238, M\'exico
         \and
             Instituto de Astrof\'{\i}sica de Canarias, V\'{\i}a L\'actea s/n, E38200 La Laguna, Tenerife, Canary Islands, Spain
         \and
             Astronomical Observatory of the Autonomous Region of the Aosta Valley,
             Loc.\ Lignan 39, 11020 Nus (AO), Italy
             }

   \date{Submitted N/A; Accepted N/A; compiled \today}

\abstract{ The TASTE project is searching for low-mass planets with
  the Transit Timing Variation (TTV) technique, by gathering
  high-precision, short-cadence light curves for a selected  sample of
  transiting exoplanets.   It has been claimed that the ``hot
  Jupiter'' WASP-3b could be perturbed by a second
  planet.  Presenting eleven new light curves (secured at the IAC80
  and UDEM telescopes) and re-analyzing thirty-eight archival light
  curves in a homogeneous way, we show that new data do not confirm
  the previously claimed TTV signal. 
  However, we bring evidence that measurements are not
  consistent with a constant orbital period, though no significant
  periodicity can be detected.  Additional dynamical modeling  and
  follow-up observations  are planned to constrain the properties of
  the perturber or to put upper limits to it.  We provide a refined
  ephemeris for WASP-3b and improved orbital/physical parameters.  A
  contact eclipsing binary, serendipitously discovered among field stars,
  is reported here for the first time.}

   \authorrunning{Nascimbeni et al.}
   \titlerunning{A homogeneous study of transit time variations in WASP-3b}
   \maketitle

\section{Introduction}

Most of the extrasolar planets discovered so far  are massive, gaseous
giant planets.  The present trend is to probe smaller and smaller
masses,  with the final aim of detecting temperate ``super-Earths'' or
Earth-sized rocky planets (1-10 $M_\oplus$, 1-3 $R_\oplus$) around
solar-type stars. The signal expected from a true Earth analog
orbiting a Sun twin is extremely small:  $\sim{}80$ ppm for the
photometric transit, and $\sim{}10$ cm/s for the radial velocity (RV)
Doppler shift. In the conventional framework both measurements are
required to derive the planetary radius and mass ($R_\mathrm{p}$,
$M_\mathrm{p}$), i.e.~the basic quantities necessary to confirm the
planetary status of the transiting body, and to characterize it.
Achieving a photometric precision of $\sim 20$ ppm  (which would allow
the detection of an Earth-like planet at 4$\sigma$ level) over the
timescale of a transit is within reach of space-based telescopes only,
while a long-term $\lesssim 20$ cm/s RV accuracy  is still too
ambitious even for the most stable spectrographs, like HARPS. 

A few indirect  techniques have been developed to get estimate of
$M_\mathrm{p}$ (or upper limits to it) for a non-transiting planet
without the need of RV measurements, the most promising being the
Transit Timing Variation (TTV) method.  By monitoring a known
transiting planet by high-precision photometry, the central instant
$T_0$ of each individual transit can be estimated.  The gravitational
perturbation of a previously-unknown third body, not necessarily
transiting, can cause a significant variation of the orbital period
$P$ \citep{holman2005}. The effect is greatly increased if the
perturber is locked in a low-order mean-motion resonance with the
transiting planet \citep{agol2005}. The  Asiago Search for Transit
timing variation of Exoplanets (TASTE)  project was started in 2010 to
search for TTVs with several ground-based, medium-class facilities on
a sample of carefully selected targets \citep{nascimbeni2010}.

The TTV technique  has been exploited and already gave interesting
results  on a number of Kepler mission candidates.  Some candidate
planets in multiple systems were validated through TTV analysis
\citep{lissauer2011}. In the past few years, some authors have
claimed TTV detections also from ground-based facilities, but  none is
confirmed so far. Among these claims,  WASP-10b \citep{mac2010b},
WASP-5b \citep{fukui2010}, the intriguing case of HAT-P-13b, that was
monitored also by TASTE
\citep{pal2011,nascimbeni2011b,southworth2012}, and the subject  of
the present investigation: WASP-3b \citep{mac2010}.

WASP-3b is a typical  short period ($P\simeq 1.8468$ days) ``hot
Jupiter'' (1.31 $R_\mathrm{jup}$, 1.76 $M_\mathrm{jup}$), hosted by a
F7-8 dwarf.  It was  discovered by \citet{pollacco2008}.  Analyzing an
$O-C$ (\emph{Observed} $-$ \emph{Calculated}) diagram computed by
comparing the $T_0$  of fourteen transits with the value predicted by
a linear ephemeris,  \citet{mac2010} claimed the detection of a
sinusoidal modulation with a period of $P_\mathrm{TTV}\simeq 127.4$
days and a semi-amplitude of $\sim 0.0014$ days $\simeq 2$ min.  They
interpreted this signal as  the effect of an outer perturber,
identifying three possible orbital solutions in the range 6-15
$M_\oplus$ and $P=3.03$-3.78 days. No independent confirmation of this
claim has been published so far, though both \citet{littlefield2011}
and \citet{sada2012} discussed the consistence of their data with that
TTV modulation.

In this paper we present (Section \ref{observations})  eleven
unpublished transits of WASP-3b: six of them have been gathered at the
IAC-80 telescope and five at the UDEM 0.36m (Universidad de Monterrey,
Mexico).   We also sifted the literature in search of  all the
archival photometric data useful for a TTV study (Section
\ref{archival}).  In Sections \ref{reduction} and \ref{analysis} we
describe how both new and archival light curves,  for a total of
forty-nine transits,   were reduced and analyzed  in a homogeneous
way, with the same software tools to provide a consistent estimate of
the planetary parameters and their uncertainties. This is crucial
especially for $T_0$, whose estimate has been shown to be easily
biased by the employed analysis technique
\citep{fulton2011,southworth2012}.  Besides $T_0$, we also refined the
orbital and physical parameters of WASP-3b, and computed an updated
ephemeris (Eq. \ref{newephem}) for any forthcoming study on this
target.  In Section \ref{analysisttv} we demonstrate that the TTV
claimed by \citet{mac2010} is not supported by our analysis, and
probably due to small-sample statistics. Yet, we point out that the
revised $O-C$ diagram displays a complex, non-periodic structure and
is not compatible with a constant orbital period.   Finally, in
Section \ref{discussion} we discuss the possible origin of this TTV
signal, and show that careful dynamical modeling and additional
photometric and RV follow-up is required to confirm the hypothesis and
to constrain  the mass and period of the possible perturber(s).

\section{TASTE observations}\label{observations}

\subsection{IAC-80 observations}

We observed six transits  of WASP-3b between 2011 May 7 and Aug 2,
employing the CAMELOT camera mounted on the IAC80 telescope.  A log
summarizing dates and other quantities of interest  is reported in
Table \ref{observ}.  Individual transits are  identified by an ID code
ranging from \texttt{N1} to \texttt{N6}.  IAC80 is a 0.8m Cassegrain
reflector installed at the Teide Observatory (Tenerife, Canary
Islands) and operated by Instituto de Astrof\'{\i}sica de Canarias
(IAC). CAMELOT is a conventional imager with a $10'.4\times 10'.4$
field of view (FOV), equipped with a E2V 42-40 2048$\times$2048 CCD
detector, corresponding to a $0.304''$ pixel scale. The software clock
interrogated to save the timestamps in the image headers is
automatically synchronized with the GPS time signal.

All the observations were carried out with a standard Bessel $R$
filter and the same instrumental setup.  Windowing and $2\times2$
binning were employed to increase the duty-cycle of the photometric
series, as described in \citet{nascimbeni2010}. A $10.4'\times 3.2'$
read-out window was chosen to image the target and a previously
selected set of reference stars in a region of the detector free from
cosmetic defects. The one-amplifier readout was preferred to prevent
gain offsets between different channels.  Stars were intentionally
defocused to a FWHM of  10-13 binned pixels ($\simeq 6.0$-$8.0''$) to
avoid saturation and to minimize systematic errors due to intra-pixel
and pixel-to-pixel inhomogeneities \citep{southworth2009b}.  Exposure
time was set to 20 s (\texttt{N1-4}) or 15 s (\texttt{N5-6}),
resulting in a net cadence $\tau=19$-25 s and a $\sim 75$-$87\%$
duty-cycle, with the only exception of \texttt{N3}. On that night, due
to software problems, the images were read unbinned and in
unwindowed readout mode, decreasing both signal-to-noise ratio
(S/N) and duty-cycle.  

The weather was photometric on all nights, except for a few thin veils
during \texttt{N3}.   Our initial goal was to start the series one
hour before the first contact of the transit, and to stop one hour
after the last contact.  Nevertheless, the amount of pre-transit
photometry is only a few minutes for the \texttt{N4} and \texttt{N6}
light curves, and the \texttt{N1} series was interrupted twenty
minutes earlier because of twilight.

\onltab{1}{
\begin{table*}
\caption{Summary of the light curves of WASP-3b analyzed in this work.}
\label{observ}
\centering
\centering\scalebox{0.865}{
\begin{tabular}{lllllllllll}\hline\hline
$N$   & ID  & evening    & telescope  & band        & $N_p$   & $\tau$  & $\sigma$ & $\sigma_{120}$ & reference paper & notes \\ 
      &     & date       &            &             &         & (s)     & (mmag)   & (mmag)  &                 &       \\ \hline
250   & \texttt{G1}  & 2008/05/18 & LT-2.0m    & $(R+V)$     & 4799    &  3.0    & 3.97     & 0.63    & \citet{gibson2008} & --- \\
309   & \texttt{G2}  & 2008/09/04 & LT-2.0m    & $(R+V)$     & 3900    &  3.0    & 3.81     & 0.61    & \citet{gibson2008} & partial \\
248   & \texttt{T1}  & 2008/05/15 & FLWO-1.2m  & Sloan $i$  & 403     &  33.8   & 1.38     & 0.73    & \citet{tripathi2010} & partial \\
262   & \texttt{T2}  & 2008/06/10 & FLWO-1.2m  & Sloan $i$  & 269     &  33.6   & 1.16     & 0.61    & \citet{tripathi2010} & partial \\
268   & \texttt{T3}  & 2008/06/21 & FLWO-1.2m  & Sloan $i$  & 522     &  34.2   & 1.59     & 0.85    & \citet{tripathi2010} & --- \\
280   & \texttt{T4}  & 2008/07/13 & UH-2.2m    & Sloan $z$  & 276     &  67.5   & 0.81     & 0.61    & \citet{tripathi2010} & --- \\
444   & \texttt{T5}  & 2009/05/12 & FLWO-1.2m  & Sloan $g$  & 224     &  78.3   & 1.17     & 0.95    & \citet{tripathi2010} & --- \\
451   & \texttt{T6}  & 2009/05/25 & FLWO-1.2m  & Sloan $g$  & 238     &  74.5   & 1.08     & 0.85    & \citet{tripathi2010} & --- \\
486   & \texttt{M1}  & 2009/07/28 & Rohzen-0.6m& $R$         & 283     &  49.3   & 2.06     & 1.32    & \citet{mac2010} & --- \\
499   & \texttt{M2}  & 2009/08/21 & Rohzen-0.6m& $R$         & 200     &  72.2   & 2.36     & 1.83    & \citet{mac2010} & --- \\
506   & \texttt{M3}  & 2009/09/03 & Rohzen-0.6m& $R$         & 297     &  45.8   & 2.87     & 1.78    & \citet{mac2010} & --- \\
519   & \texttt{M4}  & 2009/09/27 & Jena-0.9m  & $R$         & 259     &  59.6   & 2.72     & 1.92    & \citet{mac2010} & --- \\
539   & \texttt{M5}  & 2009/11/03 & Jena-0.9m  & $R$         & 258     &  51.6   & 2.41     & 1.58    & \citet{mac2010} & --- \\
629   & \texttt{M6}  & 2010/04/18 & Jena-0.9m  & $R$         & 206     &  78.0   & 1.50     & 1.21    & \citet{mac2010} & --- \\
486   & \texttt{D1}  & 2009/07/28 & OAVdA-0.25m& $R$         & 353     &  46.7   & 2.30     & 1.43    & \citet{damasso2010} & --- \\
290   & \texttt{C1}  & 2008/07/31 & HRI@EPOXI  & clear       & 1157    &  53.2   & 0.89     & 0.59    & \citet{christiansen2011} & --- \\
291   & \texttt{C2}  & 2008/08/02 & HRI@EPOXI  & clear       & 958     &  51.7   & 1.01     & 0.66    & \citet{christiansen2011} & --- \\
292   & \texttt{C3}  & 2008/08/04 & HRI@EPOXI  & clear       & 985     &  51.3   & 1.03     & 0.67    & \citet{christiansen2011} & --- \\
293   & \texttt{C4}  & 2008/08/06 & HRI@EPOXI  & clear       & 1138    &  51.3   & 0.98     & 0.64    & \citet{christiansen2011} & --- \\
294   & \texttt{C5}  & 2008/08/08 & HRI@EPOXI  & clear       & 281     &  52.2   & 0.97     & 0.64    & \citet{christiansen2011} & partial \\
296   & \texttt{C6}  & 2008/08/12 & HRI@EPOXI  & clear       & 954     &  54.6   & 0.93     & 0.63    & \citet{christiansen2011} & partial \\
297   & \texttt{C7}  & 2008/08/13 & HRI@EPOXI  & clear       & 640     &  51.5   & 1.02     & 0.67    & \citet{christiansen2011} & --- \\
298   & \texttt{C8}  & 2008/08/15 & HRI@EPOXI  & clear       & 852     &  51.5   & 1.07     & 0.70    & \citet{christiansen2011} & --- \\
654   & \texttt{L1}  & 2010/06/04 & SC-$11''$  & clear       & 218     &  76.8   & 2.68     & 2.14    & \citet{littlefield2011} & --- \\
686   & \texttt{L2}  & 2010/08/02 & SC-$11''$  & clear       & 192     &  113.   & 2.38     & 2.31    & \citet{littlefield2011} & --- \\
706   & \texttt{L3}  & 2010/09/08 & SC-$11''$  & clear       & 192     &  65.5   & 2.70     & 1.99    & \citet{littlefield2011} & --- \\
488   & \texttt{Z1}  & 2009/08/01 & Weihai-1m  & $V$         & 323     &  43.4   & 7.35     & 4.42    & \citet{zhang2011} & --- \\
444   & \texttt{S1}  & 2009/05/12 & VCT-0.5m  & Sloan $z$          & 177     &  94.7   & 2.25     & 1.99    & \citet{sada2012} & --- \\ 
653   & \texttt{S2}  & 2010/06/02 & KPNO-2.0m & $J$          & 250     &  71.8   & 1.50     & 1.16    &  \citet{sada2012} & --- \\ 
653   & \texttt{S3}  & 2010/06/02 & VCT-0.5m  & Sloan $z$          & 186     &  91.1   & 2.23     & 1.94    &  \citet{sada2012} & --- \\ 
842   & \texttt{S4}  & 2011/05/17 & VCT-0.5m  & Sloan $z$          & 217     &  87.4   & 3.52     & 3.00    &  \citet{sada2012} & --- \\ \hline
194   & \texttt{E1}  & 2009/04/22 & Newton-0.2m& clear       & 194     &  74.2   & 3.36     & 2.64    & \texttt{ETD} obs. Trnka & --- \\
446   & \texttt{E2}  & 2009/05/16 &  SC-$12''$ & $I$         & 181     &  104.   & 3.18     & 2.96    & \texttt{ETD} obs. Gregorio & --- \\
653   & \texttt{E3}  & 2010/06/02 &  SC-$12''$ & $R$         & 177     &  72.7   & 2.90     & 2.26    & \texttt{ETD} obs. Shadik \& Patrick & --- \\
666   & \texttt{E4}  & 2010/06/26 & Newton-0.3m& $B$         & 266     &  61.6   & 3.69     & 2.64    & \texttt{ETD} obs. Garlitz & --- \\
699   & \texttt{E5}  & 2010/08/26 & RC-$12.5''$& $R$         & 119     &  124.   & 2.62     & 2.67    & \texttt{ETD} obs. Hose & --- \\
838   & \texttt{E6}  & 2011/05/09 & Monteboo-0.6m& $R$       & 393     &  35.3   & 2.55     & 1.38    & \texttt{ETD} obs. Janov & --- \\ 
849   & \texttt{E7}  & 2011/05/30 & RC-$12''$  & $V$         & 297     &  51.2   & 5.65     & 3.69   & \texttt{ETD} obs. Dvorak & --- \\ \hline 
837   & \texttt{U1}  & 2008/09/03 & UDEM-0.36m & $I$         & 467     &  33.7   & 4.27     & 2.26    & this work & --- \\
844   & \texttt{U2}  & 2009/06/05 & UDEM-0.36m & $I$         & 566     &  34.1   & 3.78     & 2.01    & this work & --- \\
864   & \texttt{U3}  & 2009/07/25 & UDEM-0.36m & Sloan $z$  & 383     &  45.1   & 4.38     & 2.68    & this work & --- \\
877   & \texttt{U4}  & 2009/08/05 & UDEM-0.36m & Sloan $z$  & 371     &  44.2   & 5.03     & 3.05    & this work & low S/N \\
884   & \texttt{U5}  & 2010/08/15 & UDEM-0.36m & $I$         & 471     &  35.0   & 3.31     & 1.79    & this work & --- \\ 
837   & \texttt{N1}  & 2011/05/07 & IAC-0.8m   & $R$         & 678     &  24.0   & 1.78     & 0.79    & this work & --- \\
844   & \texttt{N2}  & 2011/05/21 & IAC-0.8m   & $R$         & 742     &  23.1   & 1.53     & 0.67    & this work & --- \\
851   & \texttt{N3}  & 2011/06/02 & IAC-0.8m   & $R$         & 412     &  38.1   & 2.93     & 1.65    & this work & low S/N \\
864   & \texttt{N4}  & 2011/06/26 & IAC-0.8m   & $R$         & 566     &  25.1   & 1.53     & 0.70    & this work & --- \\
877   & \texttt{N5}  & 2011/07/20 & IAC-0.8m   & $R$         & 785     &  20.1   & 2.12     & 0.87    & this work & --- \\
884   & \texttt{N6}  & 2011/08/02 & IAC-0.8m   & $R$         & 694     &  19.9   & 2.34     & 0.95    & this work & red noise \\ 
\hline
\end{tabular}}
\tablefoot{The columns give: the transit epoch $N$ assuming $T=T_0+NP$ and the original ephemeris (Eq. \ref{pollacco}) from 
\citet{pollacco2008}, the ID code, the ``evening date'' of the observation, the filter employed, the number of unbinned 
data points, the average net cadence $\tau$ in seconds, the photometric scatter $\sigma$ measured as the $68.27^\mathrm{th}$ percentile
of the residuals from the best-fit model, the normalized photometric scatter $\sigma_{120}=\sigma\sqrt{\tau/120^\mathrm{s}}$, 
the reference paper or database, and comments.}
\end{table*}
}

\subsection{UDEM observations}

We observed five transits of WASP-3b between 2008 Sep 3 and 2010 Aug
15 with the Universidad de Monterrey (UDEM) 0.36-m reflector. UDEM is
a small private college observatory having Minor Planet Center Code
720 located in the suburbs of Monterrey, M\'exico. The data were
acquired using standard Bessel $I$- and Sloan $z$-band filters with a
1280$\times$1024 pixel CCD camera at $1.0''$ pixel scale, resulting in
a field-of-view of $\sim21'.3\times 17'.1$. The observations were
slightly defocused to improve the photometric precision and to avoid
saturation.  On-axis guiding was used to maintain pointing
stability. Exposure times were set to 30 s for $I$ and 40 s for
$z$. All images were binned $2\times 2$ to facilitate rapid readout
($\sim 3$ s).  Each observing session lasted about 4.5 hours in order
to accommodate the transit event and also to cover about one hour
before ingress and one hour after egress. The computer clock was reset
to UTC via Internet at the beginning of every observing session to the
nearest second.

\section{Archival light curves}\label{archival}

\begin{figure*}[!p]
\centering \includegraphics[width=16.5cm]{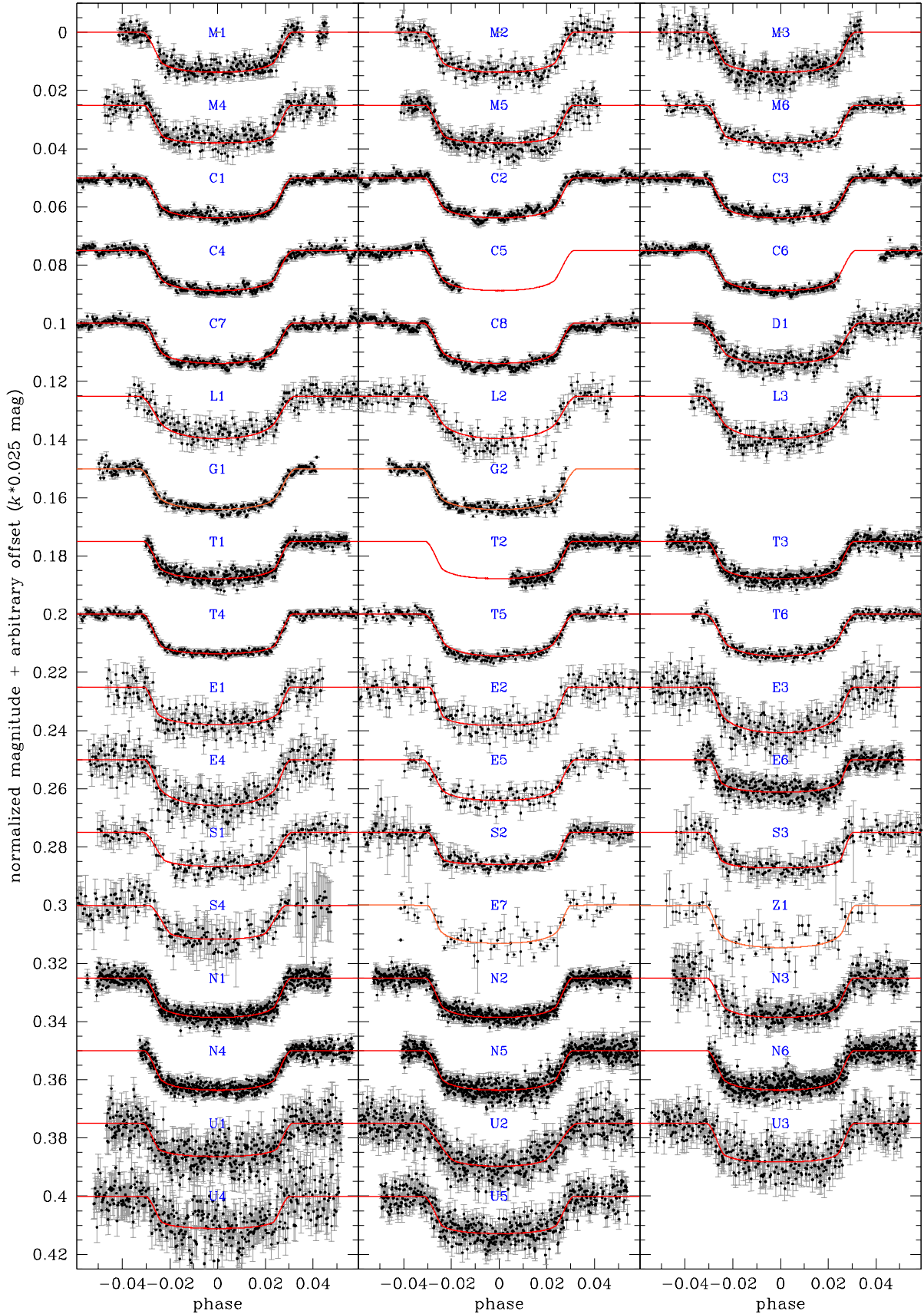}
\caption{Light curves of WASP-3b analyzed in this Paper. The
  \texttt{ID} code of each transit matches the corresponding entry in
  Table \ref{observ}.  Data points are plotted with the  original
  cadence, except for \texttt{G1}, \texttt{G2} and \texttt{Z1},
  \texttt{E7} that are binned  respectively on 30 s and 300 s
  intervals for clarity.  The red line is the best-fit model found by
  JKTEBOP. Transits have been offset in magnitude by integer multiples
  of 0.025.}
\label{lcs1}
\end{figure*}

All the available archival data were searched for any light curve
useful for a TTV analysis, i.e.~complete transits of WASP-3b with a
suitable S/N.  Partial transits were rejected, except for a few cases
with high S/N and without obvious presence of ``red noise'' after a
visual inspection. We define red noise, following  \citet{pont2006},
as correlated noise having covariance between data points on
time-scales of the same order of the duration of our transit
signal. The transit itself must be modeled and removed from a light
curve in order to study the statistical  properties of its noise
content.  Thus, first we analyzed the whole sample of light  curves in
Section \ref{analysisfit},  then we discuss how to select a
high-quality subset for furter analysis in Section \ref{analysisttv}
(``\texttt{ALL}'' vs.~``\texttt{SELECTED}'').

Hereafter we refer to each light curve with the corresponding
alphanumeric ID code  reported in Table \ref{observ}.  Overall,
thirty-eight archival light curves were (re-)analyzed in the present
study:

\begin{itemize}

\item\citet{gibson2008}: two transits in 2008 (\texttt{G1-2}) observed
  with the RISE camera mounted at the 2.0 m Liverpool Telescope. Both
  nights were photometric. The second series (\texttt{G2}) ended just
  before the last contact. The RISE camera filter has  a custom
  wideband 500-700nm,  approximately covering both the Johnson $R$ and
  $V$ passbands.

\item\citet{tripathi2010}: five transits in 2008/2009 (\texttt{T1-5})
  from two different telescopes (Fred Lawrence Whipple Observatory FLWO-1.2m
  and University of Hawaii UH-2.2m) and in three
  Sloan passbands ($g$, $i$, $z$), gathered to complement their
  high-precision RV measurements. The first two FLWO $i$ transits
  (\texttt{T1-2})  are partial, the first one missing the first
  contact by only few minutes.

\item\citet{damasso2010}: one complete transit observed in 2009
  (\texttt{D1}), with a  commercial 0.25 m reflector in a nonstandard
  $R$ band, as a part of the feasibility study for a project dedicated to
  the search for transiting rocky planets around M dwarfs.

\item\citet{mac2010}: six complete transits in the $R$ band from two
  different telescopes (\texttt{M1-3} at Rohzen-0.6m, \texttt{M4-6} at
  Jena-0.9m).  The first five observations (\texttt{M1-5}) were
  carried out on focused images, a practice resulting in a possible
  increase of the content of red noise.

\item\citet{christiansen2011}: six complete (\texttt{C1-4},
  \texttt{C7-8}) and  two partial (\texttt{C5-6}) transits, extracted
  from a  18-day photometric series secured by the High Resolution
  Imager (HRI) mounted on the spacecraft EPOXI, as part of the EPOCh
  project. No filter was employed.

\item\citet{littlefield2011}: three transits observed with a $11''$
  Schmidt-Cassegrain reflector and without filter
  (\texttt{L1-3}). These three light curves correspond to the  first,
  third, and fifth one presented in the original paper. The S/N of the
  other  curves is too low for the inclusion in the present study.

\item\citet{zhang2011}: one 2009 transit in the $V$ band from the
  Weihai-1.0m telescope  (\texttt{Z1}).

\item\citet{sada2012}: four transits secured in 2009-2011 by
  different telescopes:  \texttt{S1} and \texttt{S3-4} at the 0.5m
  Visitor Center telescope at Kitt Peak National Observatory (KPNO),
  in the Sloan $z$ band; \texttt{S2} at the KPNO 2.1m reflector in
  the $J$ band, exploiting the FLAMINGO infrared imager.

\end{itemize}

Other six light curves from various amateur observers (\texttt{E1-7}) were visually
selected and downloaded from the \texttt{ETD  database\footnote{
\texttt{http://var2.astro.cz/ETD} \citep{poddany2010}}}.

The first two follow-up light curves of WASP-3b were published by
\citet{pollacco2008}: one transit observed at the IAC80  (alternating
$V$ and $I$ filters) and one at the Keele-0.6m ($R$ band), both in
2007.  Unfortunately, the original light curves are no more
retrievable and we did not include them in this study.  A $T_0$ data
point is determined by the first term of their published ephemeris:
\begin{equation}\label{pollacco}
T_0 (\mathrm{BJD}_\mathrm{UTC}) =
2454143.8503^{+0.0004}_{-0.0003}+N\cdot 1.846
834^{+0.000002}_{-0.000002}
\end{equation}
One should be careful, however, as this $T_0$  (corresponding to 2007
Feb 12) is not an independent measurement of  a single transit
event. Instead, it comes from an ensemble analysis  of SuperWASP,
IAC80 and Keele-0.6m data. For this reason we adopted $T_0$ from
Eq.~(\ref{pollacco}) as our first data point  in the determination of
our new ephemeris, but not in our subsequent TTV analysis  (Section
\ref{analysisttv}).
The epoch $N$ of each reported observation in our set  is defined
according to the ephemeris in Eq.~(\ref{pollacco}), i.e.~as the
number of transits elapsed since 2007 Feb 12.   

All timestamps were converted to BJD(TDB),  i~e.~based on Barycentric
Dynamical Time following the prescription by \citet{eastman2010}.
Each light curve was calibrated in time by identifying the time
standard  reported in the data headers and (when clarifications were
necessary) contacting the authors.  Generally speaking, even when the
authors report a reliable synchronization source  (GPS, NTP, etc.) for
their data, it is impossible to carry out an external confirmation of
that. The only exception, discussed in Section \ref{discussion},  is
when two or more independent observations of the same event were
performed. Though we did not find any reason to doubt about the
accuracy of the absolute time calibration of the above-mentioned data,
it is worth noting that  most of these observations were not performed
with the specific goal of a TTV analysis. Thus, the precision
achievable on $T_0$ could be limited by sub-optimal choices about the
instrumental setup.  For instance, it is widely known that a ``large''
(in our case, $\tau\gtrsim 60$ s) exposure time is one of those
limiting factors \citep{kipping2010}.

When required, light curves in flux units were converted to magnitudes
and normalized to zero by fitting a low-order polynomial function to
the off-transit data points.  In a few cases, the tabulated
photometric errors are underestimated up to about 50\%, as confirmed
by the reduced $\chi^2_\mathrm{r}\gg1$ ($\chi^2$ being defined as the
$\chi^2$ divided by the number of degrees of freedom of the fit).
This is not unusual in high-precision photometry, due  to effects
which are not accounted for by standard noise models
\citep{howell2006}: poorly-modeled scintillation, high-frequency
systematics mimicking random errors, stellar microvariability for both
target and comparison stars, etc.  We dealt with this  by rescaling
the errors by a factor of $\sqrt{\chi^2_\mathrm{r}}$, following a
common practice  \citep{winn2007,gibson2009}.   When the photometric
errors were not published, the error was assumed to be constant and
equal to the scatter $\sigma$ of the off-transit polynomial-corrected
curve. We define the scatter as the $68.27^\mathrm{th}$ percentile  of
the residual distribution from the median, after a 5$\sigma$ iterative
clipping. This measurement is much more robust against outliers than
the classical RMS.

\section{Data reduction}\label{reduction}

\subsection{IAC80 photometry}

The six IAC80 light curves (\texttt{N1-6}) were reduced with the
STARSKY photometric pipeline, presented in
\citet{nascimbeni2010,nascimbeni2011b} but here upgraded with some
improvements. The present version v1.1.002 adopts a new, fully
empirical weighting scheme to carry out differential photometry.
Reference stars were previously weighted by the amount of scatter
measured on their light curves after being registered to the total
reference magnitude $m_i$ (that is, the weighted mean of the
instrumental magnitudes of all the comparison stars;
\citealt{broeg2005}). That was an iterative process.  Instead, the
updated version first extracts the off-transit part of the series,
then constructs a set of ``intermediate'' light curves of the
reference stars by subtracting  the magnitude of each of them to the
off-transit magnitude of the target. Ideally,  those curves should be
flat and their RMS should be equal to the quadratic sum
$\sqrt{\tilde\sigma^2+\sigma^2_\mathrm{t}}$, being $\sigma_\mathrm{t}$
the theoretical photometric noise expected on the target  (calculated
as in \citealt{nascimbeni2010})  and $\tilde\sigma$  the intrinsic
noise  of the reference star, defined as at the end of Section
\ref{archival}.  We therefore estimated the latter as $\tilde\sigma =
\sqrt{\sigma^2-\sigma^2_\mathrm{t}}$.   The individual weights for a
given comparison star are then assumed to be
$1/\sqrt{\tilde\sigma^2}$.  The output of this weighting algorithm is
checked during each run against two other weighting schemes: 1) equal
weights, i.e.~unweighted, and 2) using weights derived  from the
expected theoretical noise computed for the reference stars.  The
\texttt{DSYS} and \texttt{PSYS} parameters (as defined in
\citealt{nascimbeni2011b}) allow us to diagnose ``bad''  reference
stars, and to set their weights to zero. For all the
\texttt{N1}-\texttt{N6} transits, the same set of eleven reference
stars was  employed, for consistency reasons.  All of them show no
sign of variability or higher-than-expected scatter.

A second improvement to STARSKY is a new algorithm developed to deal
with light curves  having red noise caused by veils, trails or thin
clouds. This happen when the cloud possess a structure at angular
scales on the same order of the FOV,  and/or it is moving fast.  Even
differential photometry can be affected by these events,  as the
change of transparency can affect the target and the  reference stars
by a different amount. Of course, this systematic effect is correlated
on with the rapidity of transparency changes.  and this correlation
can be exploited to discard the affected frames. A quantity that we
called ``numerical derivative of the absolute flux'' (\texttt{NDAF})
is evaluated for each  frame $i$ of the series:
\begin{equation}
\texttt{NDAF}=\frac{1}{F_i}\,\frac{F_{(i+1)}-F_{(i-1)}}{t_{(i+1)}-t_{(i-1)}}
\end{equation}
being $F_i$ the weighted instrumental flux of the reference stars
($F_i=10^{-0.4m_i}$, $m_i$ defined as above), and $t_i$ the JD time at
the mid-exposure.  When \texttt{NDAF} deviates more than $4\sigma$
from its average along the series, the frame is discarded.  The first
and last frame are ignored by our algorithm.

STARSKY outputs light curves extracted from a set of different
photometric apertures. These curves are then detrended by a routine
which searches for linear and polynomial correlations between the
off-transit flux and several combinations of  external parameters such
as: the position of the star on the detector, the airmass,  the FWHM
of the stellar profiles, the mean sky level, the reference flux $F_i$,
and time.  In the following, we describe  the reason to search for a
linear correlation between  differential flux and airmass $X$.  Let's
consider  the simple case of two stars (target ``$\mathrm{t}$'' and
reference ``$\mathrm{r}$'') having out-of-atmosphere magnitudes
$m_\mathrm{t}$ and $m_\mathrm{r}$ and  color $\xi_\mathrm{t}$ and
$\xi_\mathrm{r}$, respectively.  Given the extinction coefficient $k$
and the extinction color term $k'$, the measured differential
magnitude $\Delta m$ is the difference between the observed
magnitudes:
\begin{equation} 
\Delta m = (m_\mathrm{t} + kX + k'X\xi_\mathrm{t}) - (m_\mathrm{r} + kX + k'X\xi_\mathrm{r})
\end{equation}
By grouping the involved terms, and taking into account that $X$ is the
only quantity that shows short-term variations:
\begin{equation}
\Delta m = (m_\mathrm{t}-m_\mathrm{r}) + X \cdot k'(\xi_\mathrm{t} - \xi_\mathrm{r})
\end{equation}
It is easy to see that systematic effects due to differential
extinction on the ``true'' (intrinsic) differential magnitude 
$(m_\mathrm{t}-m_\mathrm{r})$ are linearly proportional to $X$.

Eventually, we chose the light curve with the smallest amount of
scatter $\sigma$ and the lowest level of red noise estimated from the
$\beta$ parameter as defined by \citet{winn2008}. The overall S/N of
an observation can be quantified by rescaling the unbinned photometric
$\sigma$ of the light curve (having a net cadence $\tau$)  on a
standard 120-s timescale, that is by calculating
$\sigma_{120}=\sigma\sqrt{\tau/120^\mathrm{s}}$.  With the only
exception of \texttt{N3}, the final IAC80 light curves have
$\sigma_{120}=0.67$-0.95 mmag (Table \ref{observ}, Fig.~\ref{lcs2}),
only slightly larger than that  achieved by \citet{gibson2008},
\citet{tripathi2010} and \citet{christiansen2011}, with  space-based
or much larger facilities.  \texttt{N6} shows a large amount of red
noise of unknown origin, but probably related to  color-dependent
systematics caused by variable atmospheric extinction.  For the
above-mentioned reasons, \texttt{N3} and \texttt{N6}  were employed in
the determination of our new ephemeris, but not in our TTV analysis
(Section \ref{analysis}).

\subsection{UDEM photometry}

Standard dark current subtraction and twilight sky flat-field division
process were performed  for calibration on each image of the UDEM
light curves (\texttt{U1-5}).  Aperture differential photometry was
carried out on the target star and 4-6 comparison stars of similar
magnitude ($|\Delta m|\lesssim 1.5$). The apertures used varied for each date due
to defocus and weather conditions, but they were optimized to minimize
the scatter of the resulting light curves. We found that the best
results were obtained by averaging the ratios of WASP-3b to each
comparison star. This produced smaller scatter than the method of
ratioing the target star to the sum of all the comparison stars. We
estimated the formal error for each photometric point as the standard
deviation of the ratio to the individual comparison stars, divided by
the square root of their number (error of the mean). 

After normalizing the target star to the comparison stars and
averaging, some long-term systematics as a function of time were
found. This is perhaps caused by differential extinction between the
transit and comparison stars, which generally have different and
unknown spectral types. This variation was removed by fitting a linear
airmass-dependent function to the out-of-transit baseline of the light
curve.

\section{Data analysis}\label{analysis}

\subsection{Fitting of the transit model}\label{analysisfit}

We chose to analyze all of the new and archival light curves employing
the same software tools and algorithms.  Our goal was to get
homogeneous estimate of the physical/orbital parameters of WASP-3b.
In particular we were interested for our TTV analysis in estimating
the $T_0$ of each transit and its associated error in the most
accurate way, avoiding biases due to different techniques.  We avoided
estimating $T_0$ through heuristic algorithms which  assume a symmetric
light curve, such that developed by \citet{kw56}.  The main reason is
that they do not fit for any quantity other than $T_0$.  Even more
important, they are not robust against outliers, they provide values
of $T_0$ known to be  biased \citep{kipping2010} and errors on $T_0$
known to be underestimated  \citep{pribulla2012}.

JKTEBOP\footnote{\texttt{\tiny{}http://www.astro.keele.ac.uk/\textasciitilde{}jkt/codes/jktebop.html}}
\citep{southworth2004} is a code which models the light curve of a
binary system by assuming both components as biaxial ellipsoids and
performing a numerical integration in concentric annuli over the
surface of each component.  JKTEBOP version 25 was run to fit a model
light curve over our data and to derive the four main photometric
parameters of the transit: the orbital inclination $i$, the ratio of
the fractional radii $k_\mathrm{r}=R_\mathrm{p}/R_\star$, the sum of
the fractional radii $\Sigma_\mathrm{r}=R_\mathrm{p}/a+R_\star/a$
($R_\star$ is the stellar radius, $R_\mathrm{p}$ the planetary radius,
and $a$ the orbital semi-major axis), and the mid-transit time $T_0$.
{We chose to fit $i$, $k_\mathrm{r}$, $\Sigma_\mathrm{r}$
independently for each data set, because in a perturbed system $i$
and $\Sigma_\mathrm{r}$ could change over a long timescale,  while
$k_\mathrm{r}$ is an important diagnostic of  light curve quality:
when the photometric aperture is contaminated with flux from
neighbors, the transit is diluted and  $k_\mathrm{r}$ becomes
smaller. Moreover, an independent fit allows us to highlight
correlations between $i$, $k_\mathrm{r}$, $\Sigma_\mathrm{r}$ and to
derive more reliable global results, as discussed at the end of this
Section.

We set a quadratic  law to model the limb darkening (LD) effect,
naming $u_1$ the linear term  and $u_2$ the quadratic term:
$I_\mu/I_0=1-u_1(1-\mu)-u_2(1-\mu)^2$ and $\mu = \cos \gamma$, where
$I_0$ is the surface brightness at the center of the star and $\gamma$
is the angle between a line normal to  the stellar surface and the
line of sight of the observer.  \citet{southworth2010iii}, among
others, has shown that fixing the values of both $u_1$ and $u_2$
should be avoided, as it could lead to an underestimate of the
errors. On the other hand, most light curves have a S/N too low to let
both $u_1$ and $u_2$ free, and the resulting best-fit  results can be
unphysical. We set the quadratic term $u_2$ always fixed to its
theoretical value interpolated from the tables computed by
\citet{claret2000} ($BVR_\mathrm{c}I_\mathrm{c}$ bands), and
\citet{claret2004} (Sloan $ugriz$), adopting the stellar parameters of
WASP-3 derived by \citet{pollacco2008}.  For all  light curves from
nonstandard photometry, that is unfiltered CCD photometry
(\texttt{C1-8},  \texttt{L1-3}, \texttt{E1}) or from wide-band $R+V$
photometry (\texttt{G1-2}), estimating first-guess LD coefficients  is
not trivial.  As for \texttt{G1-2}, we interpolated  $u_1=0.24$ and
$u_2=0.38$  from the tables by \citet{claret2000} by taking the
average of the values tabulated for the Johnson-$V$ and Cousin-$R$
bands, as done by \citet{gibson2008}. We did the same for
\texttt{C1-8},  \texttt{L1-3}, and \texttt{E1}, assuming that the
quantum efficiency of a typical unfiltered CCD usually peaks somewhere
in between those two bands.

Then one of the three following procedures was applied:
\begin{enumerate}

\item On the data sets with a high overall $S/N$ and with two or more
  transits gathered with the same instrument and filter
  (\texttt{G1-2}; \texttt{T1-3}; \texttt{T5-T6}; \texttt{M1-3};
  \texttt{M4-5}; \texttt{C1-8}; \texttt{L1-3}; \texttt{N1-5}) we first
  fitted a model with free $i$, $k_\mathrm{r}$, $\Sigma_\mathrm{r}$,
  $T_0$ (and $u_1$ fixed at its theoretical value) to the ``best''
  individual light curves to get a preliminary estimate of their
  $T_0$.  For ``best'' we mean complete transits with high S/N:
  our choice is summarized in Table \ref{param}, fourth column.  Then we
  phased all those curves setting $T_0=0$.  The free parameters $i$,
  $k_\mathrm{r}$, $\Sigma_\mathrm{r}$, $u_1$ were fitted again on the
  stacked light curve, in order to get a high-S/N ``reference'' model
  of the transit by integrating the information contained in the whole
  set.  We fixed $i$, $k_\mathrm{r}$, $\Sigma_\mathrm{r}$, $u_1$ to
  their best-fit values,  and fitted
  a model with only $T_0$ as free parameter on all the  individual
  transits, including the low-S/N or partial ones. 

\item On \texttt{T4}, a high-S/N but single light curve, we carried
  out one simultaneous fit with $i$, $k_\mathrm{r}$,
  $\Sigma_\mathrm{r}$, $T_0$, and $u_1$ as  free parameters.

\item In all other cases, the data quality did not allow us to constrain
  $u_1$ to values with physical meaning, thus $u_1$ was fixed to its
  theoretical value along with $u_2$. Each transit was then fitted
  individually to get $i$, $k_\mathrm{r}$, $\Sigma_\mathrm{r}$, and
  $T_0$.
\end{enumerate}

The construction of the reference model from the IAC80 best data set
(\texttt{N1-2}, \texttt{N4-5} light curves) is summarized in (as an
example) Fig.~\ref{lcs2}  It is worth noting that on average the
individual light curves have $\sigma_{120}=0.75$ mmag, which decreases
to $\sigma_{120}=0.39$ mmag on the stacked data points. The expected
noise from Poissonian statistics is $0.75/\sqrt{4}=0.375$ mmag. By
taking also into account that our observations do not cover always the
same orbital phases, we are confident that the level of red noise in
the IAC80  photometry is very low.

As the formal errors derived by least squares techniques are known to
be underestimated in presence of correlated noise,  we took advantage
of two techniques implemented in JKTEBOP to estimate  realistic
errors: a Monte Carlo test (MC) and a bootstrapping method based on
the cyclic permutations of the residuals (RP or ``prayer bead''
algorithm, \citealt{southworth2008}). The errors on  all parameters
obtained with the RP algorithm are on average significantly larger for
most of the archival light curves, suggesting a non-negligible amount
of red noise.  We thus adopted conservatively the RP results in our
subsequent analysis. Mean values and error bars can be estimated in
two different ways: 1) as the arithmetic mean of the RP distribution
associated to its standard error $\pm\sigma$, and 2) as the median of
the RP distribution  along with its $15.87^\mathrm{th}$ ($\sigma_-$)
and $84.13^\mathrm{th}$ percentile ($\sigma_+$). The first estimate
assumes a Gaussian distribution, while the latter is purely empirical:
they should match in absence of red noise. 

We adopted as final results the RP median and uncertainties
$\sigma_+$, $\sigma_-$ for every fitted parameter except $T_0$.  The
estimated $T_0$ will be analyzed in  Section \ref{analysisttv} with
periodogram techniques that cannot deal with asymmetric error bars,
thus for this parameter we adopted the RP means with Gaussian errors
$\pm\sigma$.  The best-fit values of $\Sigma_\mathrm{r}$,
$k_\mathrm{r}$, $i$ and $u_1$ modeled on each individual data set are
summarized in Table \ref{param}.  We show also the theoretical value
of the linear LD coefficient $u_{1,\mathrm{th}}$ as interpolated from
\citet{claret2000,claret2004}. On all subsets, $u_1$ and
$u_{1,\mathrm{th}}$ are in agreement within $\sim 1\sigma$.  This
holds even for \texttt{C1-8}, \texttt{L1-3}, and \texttt{G1},
demonstrating that assumptions previously made on nonstandard
``clear''  or $R+V$ photometry are reasonable.

\begin{figure*}[!t]
\centering
\includegraphics[width=18cm]{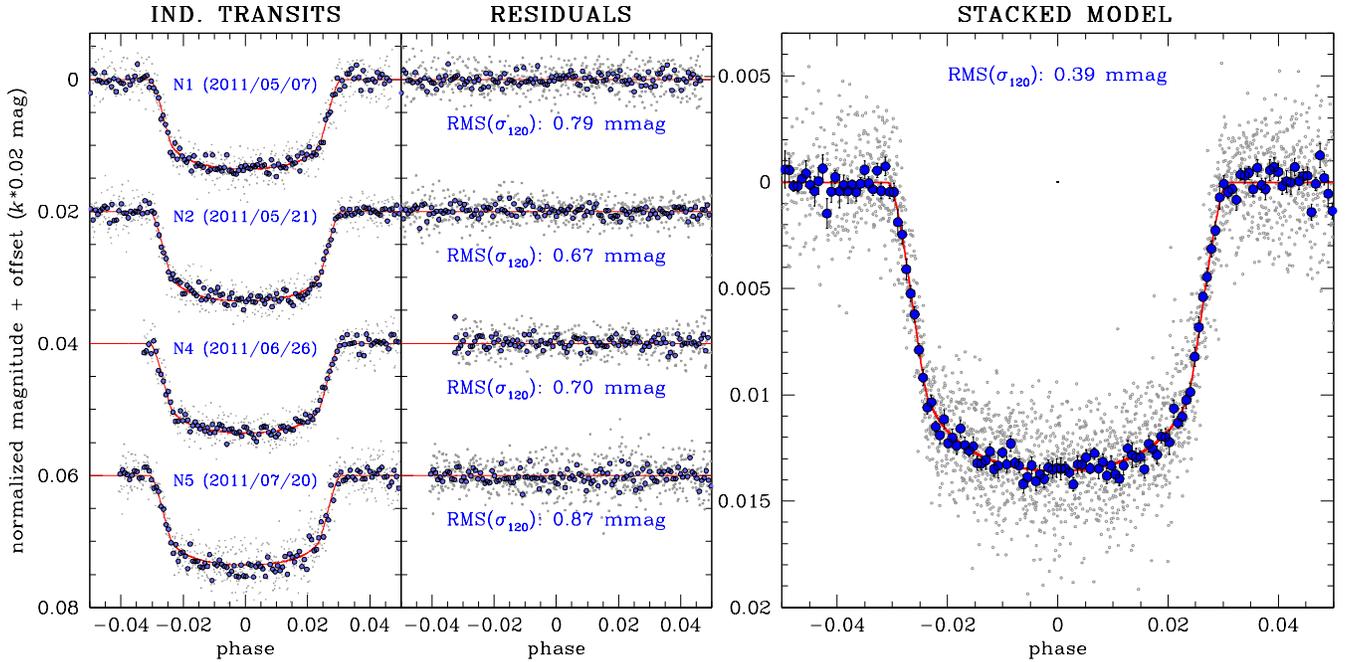}
\caption{Construction of the best-fit model from the four best light
  curves of WASP-3b observed at IAC80 (on 2011 May 7, May 21, Jun 26,
  Jul 20).  The ID\# of each transit (\texttt{N1}, \texttt{N2},
  \texttt{N4}, \texttt{N5}) matches the corresponding entry in Table
  \ref{observ}.  Small gray dots represent the data points with the
  original cadence, while blue circles are binned on 120 s
  intervals. The red line is the best-fit from JKTEBOP (Table
  \ref{param}).  \emph{Left panel:} individual light curves. Transits
  have been offset in magnitude by a multiple of 0.02 for clarity.
  \emph{Middle panel:} Residuals from the best-fit model. The reported
  scatter is evaluated  on the binned points as the $68.27^\mathrm{th}$ percentile
  from the median value.  \emph{Right panel:} Stack of all four
  IAC80 light curves with the best-fit model superimposed. The
  derived parameters are quoted in Table \ref{param} }
\label{lcs2}
\end{figure*}

In an unperturbed system, the transit parameters $\Sigma_\mathrm{r}$,
$k_\mathrm{r}$, and $i$ are purely  geometrical and should not depend
on wavelength or observing technique.  Even if the system is suspected
of being perturbed, one can check the long-term consistency of
$\Sigma_\mathrm{r}$, $k_\mathrm{r}$, and $i$ by comparing the best-fit
values estimated for independent data sets (Table \ref{param}).  It is
then possible to integrate all the extracted information to obtain
final quantities of higher precision.  We computed the weighted means
of all  subset estimate of  $\Sigma_\mathrm{r}$, $k_\mathrm{r}$, and
$i$ listed in Table \ref{param}, obtaining the result shown in the
last but one line of the same table ($\langle$weighted
mean$\rangle_1$).  As pointed out from previous works  (e.g.,
\citealt{southworth2008}), these three quantities are  correlated with
each other, as it  becomes evident by plotting their individual
estimate on planes projected from the three-dimensional parameter
space ($\Sigma_\mathrm{r}$, $k_\mathrm{r}$, $i$)
(Fig.~\ref{corr}). Apart from this, the consistency  among all
measurements is assessed within the error bars, and we conclude that
no variation of $\Sigma_\mathrm{r}$, $k_\mathrm{r}$, or $i$ is
detectable over the timescale covered by our data.  The data point
extracted from subset \texttt{T1,T3} appears  as the only probable
outlier in two correlation plots out of three.  We re-evaluated the
weighted mean after removing that point ($\langle$weighted
mean$\rangle_2$ in Table \ref{param}). The resulting averages are
quite similar to those evaluated without removing the outlier, but the
uncertainty on $k_\mathrm{r}$ is smaller.  We adopt the second mean as
final estimate:
%
%
%
%
\begin{eqnarray}
\label{newparam}
\Sigma_\mathrm{r} &=& 0.2187 \pm 0.0098 \\ 
k_\mathrm{r}      &=& 0.1058 \pm 0.0012 \\ 
i\;              &=& 84^\circ.12 \pm 0.82
\end{eqnarray}
More in general, results obtained from subsets  \texttt{T1-3} (Sloan
$i$) and \texttt{T5-6} (Sloan $g$)  are slightly but significantly in
disagreement with each other, a fact already noted by
\citet{tripathi2010} and then attributed to  the presence of residual
red noise.

\begin{figure*}[!p]
\centering
\includegraphics[height=12.0cm]{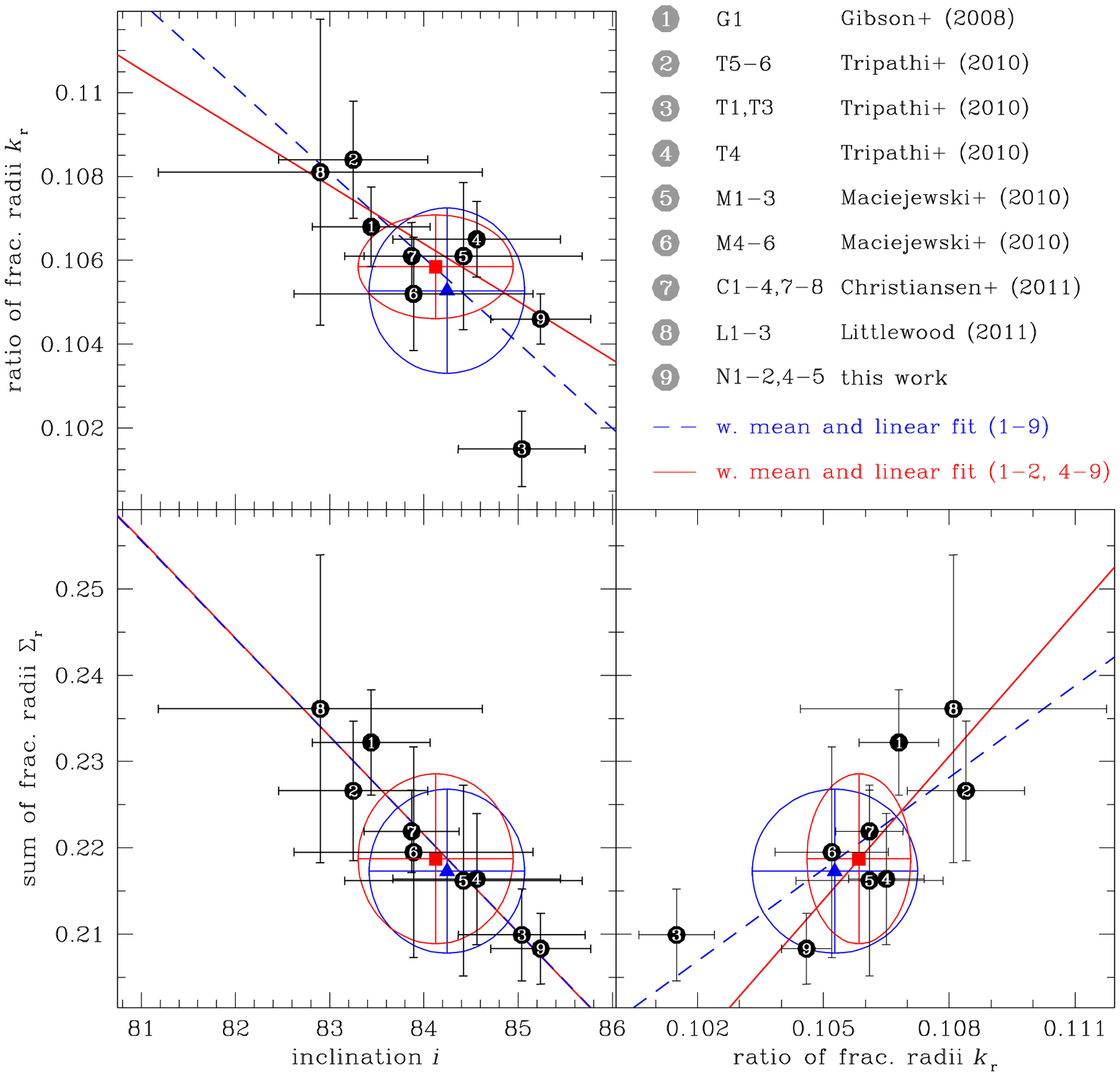}
\caption{Geometrical parameters of WASP-3b estimated from individual
  data (sub-)sets (Table \ref{param}) plotted as black labeled circles
  in their two-dimensional parameter space,  to highlight the sizeable
  correlation between $\Sigma_\mathrm{r}$, $k_\mathrm{r}$, and $i$.
  The blue dashed line in each plot is a weighted linear fit of all
  points; the blue triangles and the associated error ellipse
  correspond to the weighted mean       $\pm1\sigma$ computed on the
  marginal distribution of $\Sigma_\mathrm{r}$, $k_\mathrm{r}$, and
  $i$   (last but one row of Table \ref{param}). The red continuous
  line and squares are computed in the same way as the blue ones, but
  after removing the outlier labeled ``3'' from the set
  (\texttt{T1,T3}; last row of Table \ref{param}).}
\label{corr}
\end{figure*}

\begin{table*}[!p]
\caption{Orbital/physical parameters of WASP-3b estimated from
  individual data (sub-)sets.}
\label{param}
\centering
\centering\scalebox{0.95}{
\setlength{\extrarowheight}{6pt} 
\begin{tabular}{llllllllll}\hline\hline
reference paper & $N_\textrm{t}$ & telescope & \texttt{ID} & band  & $\Sigma_\mathrm{r}$ & $k_\mathrm{r}$ & $i$ & $u_1$ & $u_{1\mathrm{,th}}$ \\ \hline
\citet{gibson2008}& 1  &  LT-2.0m  & \texttt{G1} &  $(R+V)$    & $0.2322^{+0.0058}_{-0.0064}$   & $0.1068^{+0.0008}_{-0.0011}$ & $83^\circ.44^{+0.66}_{-0.59}$ & $0.304^{+0.069}_{-0.073}$ & 0.28 \\
\citet{tripathi2010}& 2 & FLWO-1.2m & \texttt{T5-6} & Sloan $g$    & $0.2266^{+0.0079}_{-0.0083}$   & $0.1084^{+0.0014}_{-0.0014}$ & $83^\circ.25^{+0.82}_{-0.76}$ & $0.380^{+0.086}_{-0.094}$ & 0.42 \\
\citet{tripathi2010}& 2 & FLWO-1.2m & \texttt{T1,T3} & Sloan $i$    & $0.2099^{+0.0052}_{-0.0055}$   & $0.1015^{+0.0009}_{-0.0009}$ & $85^\circ.04^{+0.69}_{-0.66}$ & $0.249^{+0.040}_{-0.057}$ & 0.18 \\
\citet{tripathi2010}& 1 & UH-2.2m & \texttt{T4} & Sloan $z$    & $0.2164^{+0.0074}_{-0.0078}$   & $0.1065^{+0.0007}_{-0.0011}$ & $84^\circ.56^{+0.96}_{-0.82}$ & $0.141^{+0.056}_{-0.054}$ & 0.14 \\
\citet{mac2010}    & 3 & Rohzen-0.6m  & \texttt{M1-3} & $R$   & $0.2162^{+0.0112}_{-0.0109}$   & $0.1061^{+0.0014}_{-0.0021}$ & $84^\circ.42^{+1.37}_{-1.15}$ & $0.221^{+0.130}_{-0.146}$ & 0.24 \\
\citet{mac2010}    & 3 & Jena-0.9m    & \texttt{M4-6} & $R$   & $0.2195^{+0.0122}_{-0.0122}$   & $0.1052^{+0.0014}_{-0.0013}$ & $83^\circ.89^{+1.31}_{-1.23}$ & $0.095^{+0.128}_{-0.129}$ & 0.24 \\
\citet{christiansen2011} & 6 & HRI@EPOXI    & \texttt{C1-4,C7-8} & clear & $0.2219^{+0.0049}_{-0.0047}$   & $0.1061^{+0.0008}_{-0.0008}$ & $83^\circ.87^{+0.51}_{-0.50}$ & $0.270^{+0.046}_{-0.054}$ & 0.28 \\
\citet{littlefield2011}  & 3 &  SC-11''  & \texttt{L1-3} & clear & $0.2361^{+0.0205}_{-0.0152}$   & $0.1081^{+0.0033}_{-0.0040}$ & $82^\circ.90^{+1.53}_{-1.91}$ & $0.455^{+0.158}_{-0.183}$ & 0.28 \\
this work & 4  & IAC80   & \texttt{N1-2,N4-5} & $R$ & $0.2083^{+0.0040}_{-0.0042}$   & $0.1046^{+0.0006}_{-0.0006}$ & $85^\circ.24^{+0.56}_{-0.50}$ & $0.247^{+0.029}_{-0.028}$ & 0.24 \\ 
$\langle$weighted mean$\rangle_1$ & -- & -- & -- & -- & $0.2173 ^{+0.0095}_{-0.0095}$ & $0.1053 ^{+0.0019}_{-0.0019}$ & $84^\circ.24 ^{+0.82}_{-0.82}$ & -- & -- \\
$\langle$weighted mean$\rangle_2$ & -- & -- & -- & -- & $0.2187 ^{+0.0098}_{-0.0098}$ & $0.1058 ^{+0.0012}_{-0.0012}$ & $84^\circ.12 ^{+0.82}_{-0.82}$ & -- & -- \\
\hline\hline
\end{tabular}}
\tablefoot{The columns give: the reference paper, the number of light
  curves fitted, the telescope, the ID code of the transits, the
  filter employed,  the fitted sum of the fractional radii
  ($\Sigma_\mathrm{r}=R_\mathrm{p}/a + R_\star/a$), the fitted ratio of the
  fractional radii  ($k_\mathrm{r}=R_\mathrm{p}/R_\star$), the fitted
  inclination $i$, the fitted linear limb darkening (LD) coefficient $u_1$,
  and the theoretical linear LD coefficient $u_{1,\mathrm{th}}$ interpolated
  from the \citet{claret2000,claret2004} tables. The
  quoted error bars are derived from the JKTEBOP RP algorithm.
  The last two rows show the weighted means of all the previous individual estimate (1)
  and of all of the previous with the exception of the data set \texttt{T1,T3} (see text for details).}
\end{table*}

\subsection{TTV analysis}\label{analysisttv}

The best-fit $T_0$ values for each transit, after being uniformly
converted to BJD(TDB), are shown in Table \ref{t0s} along with their
estimated  $\pm1\sigma$ uncertainties (second column). For completeness 
we also tabulated the median value of $T_0$ estimated from
the distribution of the RP residuals (third column of Table \ref{t0s})
along with its $15.87^\mathrm{th}$ ($\sigma_-$) and
$84.13^\mathrm{th}$ percentile ($\sigma_+$).

\onltab{3}{
\begin{table*}
\caption{Central instants of WASP-3b transits estimated from all the individual light curves.}
\label{t0s}
\centering\scalebox{0.8}{ 
\begin{tabular}{llllrrlll}\hline\hline
$N$ & $T_0$ (mean), BJD(TDB)    & $T_0$ (med), BJD(TDB) & $\Sigma$ & $O-C$ (s) & $\frac{(O-C)}{\sigma}$ & \texttt{ID}  & telescope  & selected? \\ \hline
(0) & $  2454143.85107 \pm  0.00040 $ & ---                                       &  1.00 & $ -0.00019 $ & $ -0.50 $ & \texttt{--} & --- & no (baseline ephemeris)\\
248 & $  2454601.86671 \pm  0.00026 $ & $  2454601.86673 ^{+  0.00025} _{-  0.00028} $ &  1.12 & $  0.00038 $ & $  1.48 $ & \texttt{T1} &FLWO-1.2m  & no (partial) \\
250 & $  2454605.56042 \pm  0.00030 $ & $  2454605.56049 ^{+  0.00020} _{-  0.00040} $ &  2.00 & $  0.00042 $ & $  1.42 $ & \texttt{G1} &LT-2.0m  & yes \\
262 & $  2454627.72317 \pm  0.00068 $ & $  2454627.72324 ^{+  0.00121} _{-  0.00015} $ &  8.07 & $  0.00115 $ & $  1.70 $ & \texttt{T2} &FLWO-1.2m  & no (partial) \\
268 & $  2454638.80399 \pm  0.00034 $ & $  2454638.80400 ^{+  0.00030} _{-  0.00038} $ &  1.27 & $  0.00096 $ & $  2.85 $ & \texttt{T3} &FLWO-1.2m  & yes \\
280 & $  2454660.96479 \pm  0.00015 $ & $  2454660.96480 ^{+  0.00015} _{-  0.00015} $ &  1.00 & $ -0.00025 $ & $ -1.68 $ & \texttt{T4} &UH-2.2m  & yes \\
290 & $  2454679.43318 \pm  0.00042 $ & $  2454679.43311 ^{+  0.00055} _{-  0.00029} $ &  1.90 & $ -0.00021 $ & $ -0.50 $ & \texttt{C1} &HRI@EPOXI  & yes \\
291 & $  2454681.27967 \pm  0.00034 $ & $  2454681.27963 ^{+  0.00038} _{-  0.00031} $ &  1.23 & $ -0.00055 $ & $ -1.63 $ & \texttt{C2} &HRI@EPOXI  & yes \\
292 & $  2454683.12798 \pm  0.00049 $ & $  2454683.12800 ^{+  0.00047} _{-  0.00051} $ &  1.09 & $  0.00092 $ & $  1.88 $ & \texttt{C3} &HRI@EPOXI  & yes \\
293 & $  2454684.97524 \pm  0.00040 $ & $  2454684.97521 ^{+  0.00042} _{-  0.00038} $ &  1.11 & $  0.00134 $ & $  3.36 $ & \texttt{C4} &HRI@EPOXI  & yes \\
294 & $  2454686.82144 \pm  0.00074 $ & $  2454686.82140 ^{+  0.00102} _{-  0.00047} $ &  2.17 & $  0.00071 $ & $  0.96 $ & \texttt{C5} &HRI@EPOXI  & no (partial)\\
296 & $  2454690.51475 \pm  0.00059 $ & $  2454690.51480 ^{+  0.00048} _{-  0.00070} $ &  1.46 & $  0.00035 $ & $  0.59 $ & \texttt{C6} &HRI@EPOXI  & no (partial)\\
297 & $  2454692.36168 \pm  0.00056 $ & $  2454692.36173 ^{+  0.00047} _{-  0.00064} $ &  1.36 & $  0.00044 $ & $  0.79 $ & \texttt{C7} &HRI@EPOXI  & yes \\
298 & $  2454694.20776 \pm  0.00083 $ & $  2454694.20770 ^{+  0.00090} _{-  0.00075} $ &  1.20 & $ -0.00031 $ & $ -0.37 $ & \texttt{C8} &HRI@EPOXI  & yes \\
308 & $  2454712.67641 \pm  0.00064 $ & $  2454712.67637 ^{+  0.00066} _{-  0.00062} $ &  1.06 & $ -0.00000 $ & $ -0.01 $ & \texttt{U1} &UDEM-0.36m & yes \\
309 & $  2454714.52368 \pm  0.00041 $ & $  2454714.52358 ^{+  0.00063} _{-  0.00020} $ &  3.15 & $  0.00042 $ & $  1.04 $ & \texttt{G2} &LT-2.0m    & no (partial)\\
433 & $  2454943.53240 \pm  0.00060 $ & $  2454943.53249 ^{+  0.00045} _{-  0.00074} $ &  1.64 & $  0.00161 $ & $  2.70 $ & \texttt{E1} &Newton-0.2m  & yes \\
444 $\star$ & $  2454963.84450 \pm  0.00081 $ & $  2454963.84453 ^{+  0.00087} _{-  0.00075} $ &  1.16 & $ -0.00146 $ & $ -1.81 $ & \texttt{T5} &FLWO-1.2m  & yes \\
444 $\star$ & $  2454963.84527 \pm  0.00118 $ & $  2454963.84541 ^{+  0.00102} _{-  0.00134} $ &  1.31 & $ -0.00069 $ & $ -0.59 $ & \texttt{S1} &VCT-0.5m  & yes \\
446 & $  2454967.53905 \pm  0.00070 $ & $  2454967.53912 ^{+  0.00061} _{-  0.00079} $ &  1.30 & $ -0.00058 $ & $ -0.84 $ & \texttt{E2} &SC-$12''$  & yes \\
451 & $  2454976.77284 \pm  0.00030 $ & $  2454976.77283 ^{+  0.00032} _{-  0.00028} $ &  1.14 & $ -0.00097 $ & $ -3.23 $ & \texttt{T6} &FLWO-1.2m  & yes \\
457 & $  2454987.85256 \pm  0.00093 $ & $  2454987.85267 ^{+  0.00071} _{-  0.00116} $ &  1.63 & $ -0.00225 $ & $ -2.43 $ & \texttt{U2} &UDEM-0.36m  & yes \\
484 & $  2455037.71878 \pm  0.00086 $ & $  2455037.71877 ^{+  0.00085} _{-  0.00086} $ &  1.01 & $ -0.00058 $ & $ -0.68 $ & \texttt{U3} &UDEM-0.36m  & yes \\
486 $\star$ & $  2455041.41172 \pm  0.00035 $ & $  2455041.41172 ^{+  0.00036} _{-  0.00035} $ &  1.03 & $ -0.00131 $ & $ -3.75 $ & \texttt{D1} &OAVdA-0.25m  & yes \\
486 $\star$ & $  2455041.41255 \pm  0.00058 $ & $  2455041.41246 ^{+  0.00053} _{-  0.00063} $ &  1.19 & $ -0.00048 $ & $ -0.83 $ & \texttt{M1} &Rohzen-0.6m  & yes \\
488 & $  2455045.10565 \pm  0.00106 $ & $  2455045.10557 ^{+  0.00112} _{-  0.00100} $ &  1.12 & $ -0.00105 $ & $ -0.99 $ & \texttt{Z1} &Weihai-1m  & no (low S/N) \\
490 & $  2455048.80065 \pm  0.00107 $ & $  2455048.80066 ^{+  0.00104} _{-  0.00110} $ &  1.06 & $  0.00027 $ & $  0.26 $ & \texttt{U4} &UDEM-0.36m  & no (red noise) \\
499 & $  2455065.42023 \pm  0.00036 $ & $  2455065.42026 ^{+  0.00038} _{-  0.00034} $ &  1.12 & $ -0.00165 $ & $ -4.60 $ & \texttt{M2} &Rohzen-0.6m  & yes \\
506 & $  2455078.34809 \pm  0.00114 $ & $  2455078.34799 ^{+  0.00133} _{-  0.00096} $ &  1.39 & $ -0.00163 $ & $ -1.44 $ & \texttt{M3} &Rohzen-0.6m  & yes \\
519 & $  2455102.36030 \pm  0.00084 $ & $  2455102.36037 ^{+  0.00061} _{-  0.00107} $ &  1.75 & $  0.00171 $ & $  2.04 $ & \texttt{M4} &Jena-0.9m  & yes \\
539 & $  2455139.29753 \pm  0.00073 $ & $  2455139.29719 ^{+  0.00111} _{-  0.00034} $ &  3.26 & $  0.00224 $ & $  3.08 $ & \texttt{M5} &Jena-0.9m  & yes \\
629 & $  2455305.51117 \pm  0.00056 $ & $  2455305.51119 ^{+  0.00049} _{-  0.00063} $ &  1.29 & $  0.00074 $ & $  1.34 $ & \texttt{M6} &Jena-0.9m  & yes \\
653 $\star$ & $  2455349.83306 \pm  0.00090 $ & $  2455349.83316 ^{+  0.00085} _{-  0.00096} $ &  1.13 & $ -0.00140 $ & $ -1.56 $ & \texttt{S3} &VCT-0.5m  & yes \\
653 $\star$ & $  2455349.83390 \pm  0.00069 $ & $  2455349.83384 ^{+  0.00069} _{-  0.00069} $ &  1.00 & $ -0.00056 $ & $ -0.81 $ & \texttt{S2} &KPNO-2.0m  & yes \\
653 $\star$ & $  2455349.83434 \pm  0.00054 $ & $  2455349.83430 ^{+  0.00061} _{-  0.00047} $ &  1.30 & $ -0.00012 $ & $ -0.22 $ & \texttt{E3} &SC-$12''$  & yes \\
654 & $  2455351.68410 \pm  0.00155 $ & $  2455351.68399 ^{+  0.00151} _{-  0.00159} $ &  1.05 & $  0.00280 $ & $  1.81 $ & \texttt{L1} &SC-11''  & no (red noise)\\
666 & $  2455373.84289 \pm  0.00045 $ & $  2455373.84285 ^{+  0.00048} _{-  0.00043} $ &  1.12 & $ -0.00042 $ & $ -0.94 $ & \texttt{E4} &Newton-0.3m  & yes \\
686 & $  2455410.78073 \pm  0.00173 $ & $  2455410.78083 ^{+  0.00148} _{-  0.00199} $ &  1.34 & $  0.00071 $ & $  0.42 $ & \texttt{L2} &SC-11''  & no (red noise)\\
693 & $  2455423.70889 \pm  0.00048 $ & $  2455423.70891 ^{+  0.00050} _{-  0.00047} $ &  1.06 & $  0.00103 $ & $  2.15 $ & \texttt{U5} &UDEM-0.36m  & yes \\
699 & $  2455434.78772 \pm  0.00051 $ & $  2455434.78775 ^{+  0.00041} _{-  0.00060} $ &  1.46 & $ -0.00114 $ & $ -2.25 $ & \texttt{E5} &RC-$12.5''$  & yes \\
706 & $  2455447.71600 \pm  0.00117 $ & $  2455447.71628 ^{+  0.00051} _{-  0.00184} $ &  3.61 & $ -0.00071 $ & $ -0.61 $ & \texttt{L3} &SC-11''  & no (red noise)\\
837 & $  2455689.65263 \pm  0.00015 $ & $  2455689.65263 ^{+  0.00014} _{-  0.00016} $ &  1.14 & $  0.00054 $ & $  3.66 $ & \texttt{N1} &IAC-0.8m  & yes \\
838 & $  2455691.49899 \pm  0.00064 $ & $  2455691.49899 ^{+  0.00071} _{-  0.00056} $ &  1.27 & $  0.00007 $ & $  0.11 $ & \texttt{E6} &Monteboo-0.6m  & yes \\
842 & $  2455698.88476 \pm  0.00160 $ & $  2455698.88463 ^{+  0.00174} _{-  0.00147} $ &  1.18 & $ -0.00149 $ & $ -0.94 $ & \texttt{S4} &VCT-0.5m  & yes \\
844 & $  2455702.58052 \pm  0.00028 $ & $  2455702.58051 ^{+  0.00029} _{-  0.00028} $ &  1.04 & $  0.00059 $ & $  2.12 $ & \texttt{N2} &IAC-0.8m  & yes \\
849 & $  2455711.81615 \pm  0.00107 $ & $  2455711.81617 ^{+  0.00094} _{-  0.00120} $ &  1.28 & $  0.00204 $ & $  1.92 $ & \texttt{E7} &RC-$12''$  & no (low S/N) \\
851 & $  2455715.50882 \pm  0.00057 $ & $  2455715.50853 ^{+  0.00051} _{-  0.00063} $ &  1.24 & $  0.00105 $ & $  1.84 $ & \texttt{N3} &IAC-0.8m  & no (low S/N) \\
864 & $  2455739.51620 \pm  0.00017 $ & $  2455739.51620 ^{+  0.00015} _{-  0.00018} $ &  1.20 & $ -0.00042 $ & $ -2.49 $ & \texttt{N4} &IAC-0.8m  & yes \\
877 & $  2455763.52511 \pm  0.00031 $ & $  2455763.52511 ^{+  0.00030} _{-  0.00032} $ &  1.07 & $ -0.00036 $ & $ -1.19 $ & \texttt{N5} &IAC-0.8m  & yes \\
884 & $  2455776.45192 \pm  0.00089 $ & $  2455776.45192 ^{+  0.00093} _{-  0.00085} $ &  1.09 & $ -0.00140 $ & $ -1.58 $ & \texttt{N6} &IAC-0.8m  & no (red noise)\\

\hline\hline
\end{tabular}}
\tablefoot{The columns give: the transit epoch $N$ assuming $T=T_0+NP$
  and the original ephemeris from  \citet{pollacco2008}, the best-fit
  value for the central instant $T_0$ of the transit and the
  associated $\sigma$, the median value of the distribution of $T_0$
  from the Residual Permutation (RP) algorithm and the associated
  errors $\sigma_+$ and $\sigma_-$, the ``skew'' parameter  $\Sigma =
  \max\{ \sigma_+,\sigma_- \} / \min \{ \sigma_+,\sigma_- \} $, the
  $O-C$ according to the new ephemeris (Eq. \ref{newephem}) in
  seconds, the $O-C$ in units of $\sigma$,  the ID code of the light
  curve, the telescope employed, and comments about whether and why
  the data point is excluded from the ``selected''
  sample. Simultaneous transits are marked with a star in the first
  column.}
\end{table*}
}

To check which light curves  within our sample are  significantly
affected by red noise, we consider two  diagnostic parameters named
$\beta$ and $\Sigma$. The first one is defined as in \citet{winn2008}:
the light curve is averaged over $M$ bins containing $N$ unbinned
points each, then $\beta$ is calculated as
\begin{equation}
\beta=\frac{\sigma_\mathrm{N}}{\tilde\sigma_\mathrm{N}}\textrm{ ,}\qquad
\tilde\sigma_\mathrm{N}=\frac{\sigma}{\sqrt{N}}\sqrt{\frac{M}{M-1}}
\end{equation}
That is, $\beta$ is the ratio between the scatter $\sigma_\mathrm{N}$
measured on a given temporal scale $\Delta t = N\tau$,  and the
expected noise $\tilde\sigma_\mathrm{N}$  estimated by rescaling the
unbinned scatter $\sigma$ assuming Gaussian  statistics.  Ideally, we
expect $\beta\simeq 1$ for  independent and random errors (i.e.~pure
``white noise''),  larger values meaning presence of red noise at time
scales $\sim\Delta t$.  The time scales around $\Delta t \simeq25$ min
are the most important ones to our purposes, because they correspond
to the duration of the ingress/egress part of the WASP-3b light
curve. As shown by \citet{doyle2004}, those are the parts having the
largest information content about $T_0$.  We chose to compute $\beta$
on a set of averaging times between  $\Delta t=20$ and 30 minutes,
then we took their arithmetic mean  as a final estimate for $\beta$. 

The second diagnostic is $\Sigma$. We mentioned above  that the
distribution of the RP residuals  around the best-fit value  should be
symmetric if the noise budget is dominated by white noise.    Hence a
``skewed'' distribution could highlight a non-negligible amount of red
noise. The opposite is not always true: short-term systematics
($\Delta t\lesssim\tau$) do not necessarily lead to skewed RP
residual distributions.  We parametrized this ``skewness'' with the
ratio $\Sigma$ between the largest and the smallest error bar $\sigma$
of a given data point:
\begin{equation}
\Sigma = \max\{ \sigma_+,\sigma_- \} / \min \{ \sigma_+,\sigma_-
\}
\end{equation}
In principle, $\Sigma \simeq 1$ for well-behaved transits, and $\Sigma
\gg 1$ for transits dominated by long-term systematics.  Table
\ref{t0s} lists $\Sigma$ for all the employed $T_0$. We found
$1.01<\Sigma<1.63$ for the eleven TASTE transits (\texttt{N1-6},
\texttt{U1-5}). Instead, a few archival data points show unusually
large values (e.g., $\Sigma=8.07$ for \texttt{T2}). We investigated
this issue by comparing the most significant $T_0$ published in the
literature  (\texttt{T1-5}; \texttt{G1-2}; \texttt{C1-8};
\texttt{M1-6}) with those derived by our reanalysis (Fig.~\ref{oc},
third panel from the top).  We concluded that the vast majority of the
published estimate agree  within the error bars with ours.  Notable
exceptions are \texttt{T1}, \texttt{T2}, and  \texttt{G2}: they all
are \emph{partial} transits, and \texttt{T2}, \texttt{G2} have  also
large $\Sigma$. This is exactly what we expected. It indeed
demonstrates that when a light curve lacks the off-transit part, its
normalization becomes tricky, and even a very small difference in the
adopted technique can lead to significantly different $T_0$
measurements.  We emphasize this conclusion because many TTV studies
employ partial transits (some of them even for the most part: among
others, \citealt{pal2011} and \citealt{fulton2011}).  While this is
fine when estimating the orbital and physical parameters of the
planet, we demonstrated that partial light curves should be included
with extreme caution in a TTV analysis. We note that, on average, our
error bars are larger than the published ones --sometimes by a factor
of two-- confirming our concern that most measurements carried out in
the past have been published with underestimated errors due to
neglected red noise. 

We considered two different samples of measurements for our TTV
analysis. The first  (``\texttt{ALL}'') includes all the 49 $T_0$
listed in Table \ref{t0s}, plus the $T_0$ from the ephemeris
Eq.~(\ref{pollacco}) from \citet{pollacco2008}. The second sample
(``\texttt{SELECTED}'') is a high-quality subset of 36 values,
selected by excluding  Eq.~(\ref{pollacco}) (it does not correspond to
an independent measurement of one single transit) and other 13 data
points using the following rejection criteria:

\begin{enumerate}
\item partial light curve: data points lacking between the first and
  the last contact;
\item large scatter: $\sigma_{30}> 1$ mmag ($\sigma_{30}$ defined  as
  the RMS of the residuals after averaging over 30-min bins);
\item red noise: $\Sigma > 2$ or $\beta > 1.2$;
\item presence of systematics during or close to the transit
  ingress/egress, as determined by visual inspection of  the residuals.
\end{enumerate}
The transits excluded from  \texttt{SELECTED} are listed in Table
\ref{t0s} along with the reason for the exclusion.

The first step to plot an $O-C$ diagram is to calculate a
``reference'' linear ephemeris to predict $T_0$ at any given epoch.
We set the new zero epoch at \texttt{N2}, i.e.~our most accurate
light curve. The \texttt{ALL} sample was employed to fit a linear
model by ordinary weighted least squares, obtaining
\begin{eqnarray}
\label{newephem}
T_0 (\mathrm{BJD}_\mathrm{TDB})  &=& 2\,455\,702.57993 \pm 0.00017 +
\nonumber \\ && + N\cdot 1.8468349 \pm 0.0000004 
\end{eqnarray}

The uncertainties have been evaluated from the covariance matrix of
the fit, and were both rescaled by $\sqrt{\chi^2_\mathrm{r}}$ to take
into account the real dispersion of the data points around our
best-fit ephemeris. In Fig.~\ref{oc} are plotted the $O-C$ diagrams
for \texttt{ALL} (first panel from top) and \texttt{SELECTED} samples
(second panel, with a smaller baseline). In both diagrams, the reduced
$\chi^2_\mathrm{r}=3.7$-4.5 suggests that the measurements are not in
full agreement with the linear ephemeris in Eq.~(\ref{newephem}).
Yet, there is no evident periodic pattern in our diagrams. We
investigated the possibility that this statistically significant
scatter is caused by a genuine TTV, either the TTV claimed by
\citet{mac2010} or a different one.

\begin{figure*}[!p]
\centering \includegraphics[width=16.1cm]{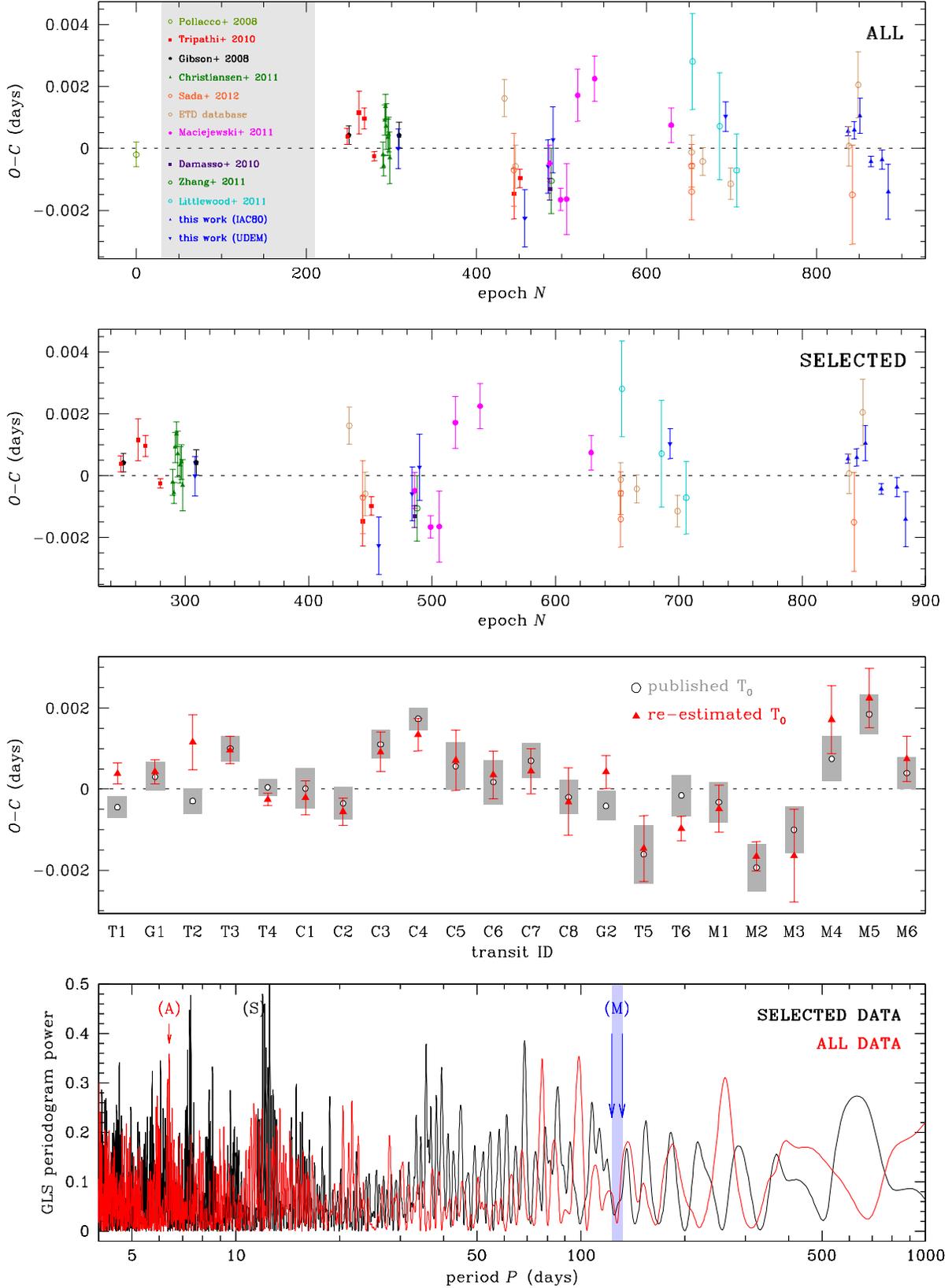}
\caption{\emph{First panel from the top:} $O-C$ diagram for all  data
  points tabulated in Table \ref{t0s}. \emph{Second panel:} same as
  above, for points selected as in the last column of Table \ref{t0s}.
  \emph{Third panel:} Comparison between the original $T_0$ published
  by the respective authors (white circles with gray error bars; Table
  \ref{observ})  and as re-estimated in this work (red triangles and
  bars), for the subset of high-precision light curves identified in
  the horizontal axis. \emph{Fourth panel} GLS periodogram for the
  complete sample (red line, highest peak at $A$) and the selected
  sample (black line, highest peak at $S$). The periodicity claimed by
  \citet{mac2010} is marked with the $M$ label.}
\label{oc}
\end{figure*}

We exploited two algorithms to search for periodic signals: the
Generalised Lomb-Scargle periodogram (GLS; \citealt{zechmeister2009})
and the ``Fast $\chi^2$'' algorithm (F$\chi^2$; \citealt{palmer2009}).
Both these techniques are able to deal with irregularly sampled data
with nonuniform weights, and minimize aliasing effects due to the
window function. In addition to this, $F\chi^2$ can also search for an
arbitrary number of harmonics.

We searched for periodic signals with GLS in the period range
$P=4$-1000 d, the lower limit being imposed by the Nyquist sampling
criterion to avoid aliasing \citep{horne1986}.  The resulting
periodograms for both samples are plotted in the fourth panel of
Fig.~\ref{oc}. In neither case a prominent peak is visible. The
\texttt{ALL} and \texttt{SELECTED} periodograms are quite similar, and
their highest peaks stand at $P(\texttt{A})\simeq 6.41$ d and
$P(\texttt{S})\simeq 11.29$ d, respectively. 
The formal false alarm probability (FAP)
 as defined by \citet{zechmeister2009} is 
0.023 (2.28$\sigma$) for the \texttt{ALL} power peak and 0.076 (1.77 $\sigma$)
for the \texttt{SELECTED} peak, i.~e.~only marginally significant. However,
the formal FAP is derived under the assumption of pure Gaussian noise,
which is not our case.
To take into account the intrinsic dispersion of
our data, we investigated whether these
peaks are statistically significant or not with a resampling
algorithm. We generated $10\,000$ synthetic $O-C$ diagrams with the
same temporal coordinates of actual data points, by randomly
scrambling the $O-C$ values at each generation. A GLS periodogram was
then evaluated on each of them with the same settings of that applied
on real data. The power of the highest peak found in the real data for
\texttt{ALL} and \texttt{SELECTED} samples lies respectively at the
$12^\mathrm{th}$  ($-1.17\sigma$) and $38^\mathrm{th}$ percentile
($-0.31\sigma$) of the distribution of the maximum-power peaks in the
synthetic,  randomly-permutated diagrams. We conclude that neither
peaks can be considered as statistically significant.  In particular,
the $P(\texttt{M})\simeq 127$ days periodicity claimed by
\citet{mac2010} is not consistent with our data. Instead of a peak, the
periodogram range  where  the $P(\texttt{M})$ peak is expected is
characterized by an extremely low  GLS power (fourth panel of
Fig.~\ref{oc}, blue region). On the other hand, tests on 
synthetic $O-C$ diagrams having the same sampling and noise properties of our
sample demonstrate that a $P=P(\texttt{M})$, $\Delta(O-C)=0.0014$ days  
signal corresponding to the \citet{mac2010} claim
would be easily detectable from our data.

We set F$\chi^2$ to search for periodicities with one or two harmonics
in the same frequency range.  Results from both samples are quite
similar to those obtained above with GLS, with non-significant power
peaks  at periods very close to those previously found in the GLS
periodogram.

\begin{figure*}[tp]
\centering \includegraphics[width=0.8\textwidth,clip=true]{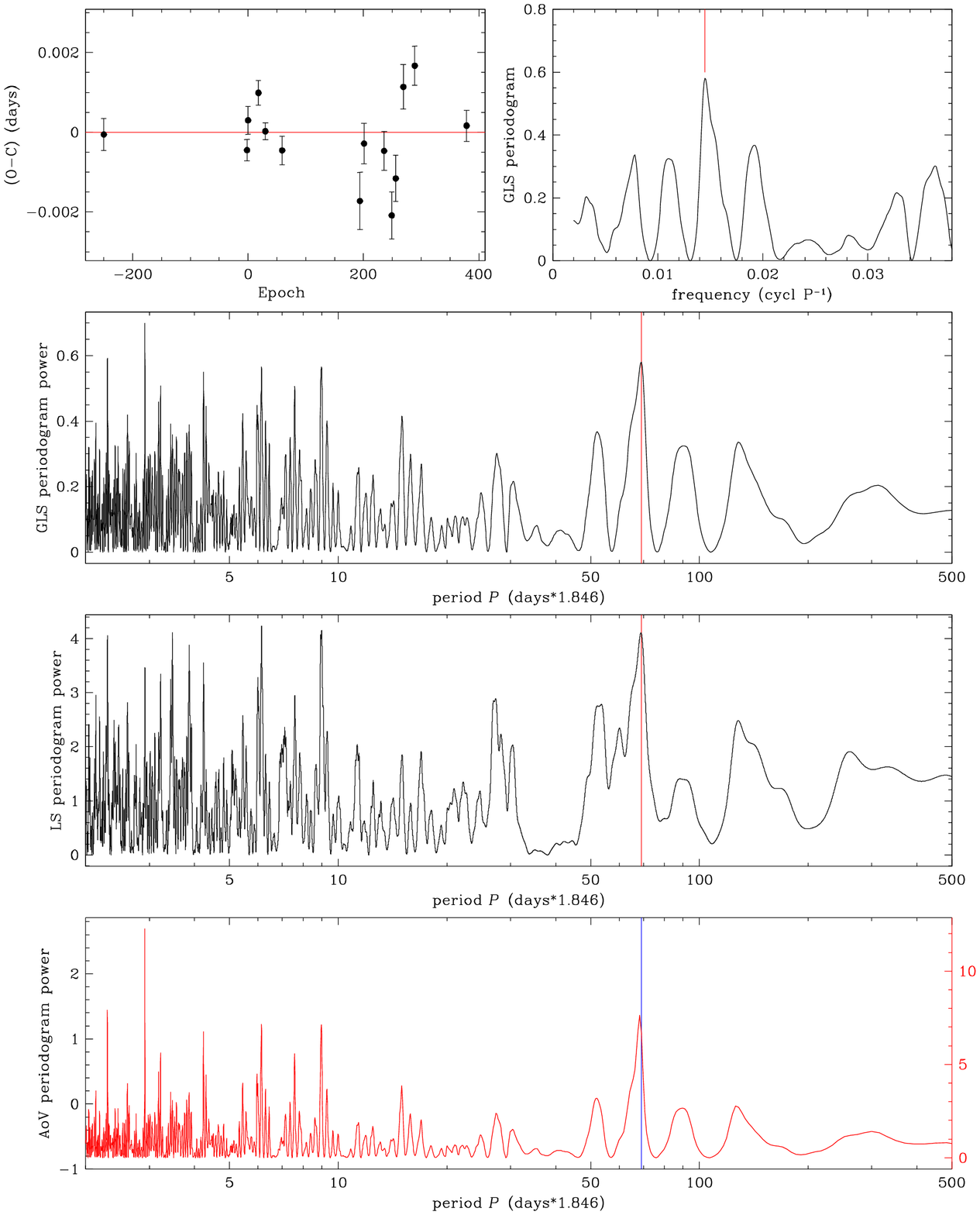}
\caption{Period analysis carried out on the same set of $O-C$ data
  points analyzed by \citet{mac2010}. \emph{Upper left panel:} $O-C$
  diagram for the selected points.  \emph{Upper right panel: } GLS
  periodogram as a function of frequency $\nu$, adopting the same
  plotting  limits of \citet{mac2010}.  \emph{Bottom panel: } GLS
  periodogram as a function of period, adopting wider limits on
  frequency according to the Nyquist criterion (see text for details).
  The red vertical line marks the peak claimed by \citet{mac2010}.}
\label{repro}
\end{figure*}

We carried out our analysis also on a subset of data points
corresponding to those analyzed by \citet{mac2010}, by employing the
same tools used for our full data set (Fig.~\ref{repro}).  When
plotted adopting the same frequency limits,  the resulting periodogram
(and the peak corresponding to the maximum power)  perfectly matches
those published by \citet{mac2010} (upper right panel of
Fig.~\ref{repro}). On the other hand, the upper limit in frequency set
by \citet{mac2010} (0.038 $P^{-1}$) is too low according to  the
Nyquist criterion. In the range  $0.038 < \nu < 0.5$ $P^{-1}$ many
other maxima are as high as the 0.0145 $P^{-1}$ peak (lower panel of
Fig.~\ref{repro}). Following a statistical test  similar to  the one
above described,  we discard  that peak as not significant, being due
to small-sample statistics.

\section{Discussion and conclusions}\label{discussion}

In the present study, we analyzed eleven unpublished light  curves of
WASP-3b and re-analyzed other thirty-eight archival light curves, all
of them with the same software tools and procedures. We derived
improved orbital and physical parameters for this target (Table
\ref{param}),  and computed a refined ephemeris (Eq.~\ref{newephem}).
All individual measurements of the central instant $T_0$ have been
compared with the new ephemeris to search for changes in the orbital
period $P$ of the transiting planet.  We concluded that available
observations of WASP-3b, spanning more than four years, are not
consistent with a linear ephemeris ($\chi^2_\mathrm{r}\simeq4$). A
possible explanation for this scatter is the presence of a perturbing
body in the WASP-3 planetary system.

It is known that the impact of red noise on high-precision transit
photometry is still not fully understood. Previous claims of TTVs have
been disproved on this basis \citep{southworth2012,fulton2011}.  Could
the observed scatter in the $O-C$ diagram of WASP-3b  be explained in
terms of  underestimated observational errors or calibration issues?
The absolute time calibration of each archival light curve cannot be
independently checked.  In principle one should trust the authors
about that.  However, we point out two main clues supporting the
auto-consistence of the overall data, and hence the hypothesis of a
genuine TTV:

\begin{itemize}
\item on three different epochs ($N=444$, 486, 653 following the
  \citealt{pollacco2008} ephemeris) multiple observations of the same
  transit are available. As they were carried out by different authors
  at different facilities, they should represent independent
  measurements of the same quantity. All these data points (marked
  with a star in the first column of Table \ref{t0s}) agree with each
  other within their $1\sigma$ error bars, suggesting that the
  uncertainties on $T_0$ are correctly evaluated by our pipeline;
\item some anomalous patterns in the $O-C$ diagrams are confirmed by
  several different data sets. For instance, nearly \emph{all}  points
  gathered in 2009 within the range $N=440$-510 lie ahead of the
  $T_0$ predicted by our baseline ephemeris ($O-C<0$). The only
  exception is \texttt{U4}, which essentially lies at $O-C\sim 0$
  within its error bar. The weighted mean of these twelve measurements
  from eight different authors  is $O-C=-0.00118\pm 0.00016$
  days. This implies a $7.2\sigma$ deviation from a constant orbital
  period. These patterns can also be detected in high-precision data
  subsets, such as our ones. Among the four best IAC80 transits, the
  first two (\texttt{N1-2}) are delayed  by $4.3\sigma$
  (i.e.~$O-C=48\pm11$ s) compared with the  prediction, while the
  second two (\texttt{N4-5}) are ahead of the ephemeris by $2.8\sigma$
  ($O-C= 35 \pm 13$ s).
\end{itemize}

The available data thus strongly suggest an intrinsic  deviation of
the actual transit times from the linear ephemeris expected from a
Keplerian two-body transiting system.  On the other hand, our analysis
rules out the periodic TTV claimed by \citet{mac2010}, and failed even
at detecting any significant periodicity in the updated $O-C$
diagram. Most TTV studies carried out in the past searched for a
periodic signal, but here we are dealing with a more complex and
non-periodic phenomenon, at least  at the time scales we sampled. What
kind of dynamical system can induce such a perturbation?

\citet{veras2011} demonstrated that some orbital configurations,
especially close to (but not exactly in) mean-motion resonances,  can
induce quasi-periodic or even chaotic TTVs. In other non-exotic
configurations, the periodicity would manifest itself only at time
scales $>10$ yr  \citep{veras2011}. Also when more than one perturber
are present and their orbital periods are not commensurable, as in the
case of our inner Solar System, the resulting TTV would be in general
aperiodic \citep{holman2005}.

If WASP-3b belongs to one of the above-mentioned cases, careful dynamical
modeling and additional follow-up is required to confirm the
hypothesis and to constrain  the mass and period of the possible
perturber(s). Photometric monitoring is still ongoing within the TASTE
project, and high-precision RV measurements are foreseen with
HARPS-N. As stressed out by \citet{meschiari2010} and
\citet{payne2011},  photometric TTVs and RVs are highly complementary
in breaking the degeneracies that are  common in the inverse dynamical
problem. 

\begin{acknowledgements}

This work was partially supported by PRIN INAF 2008 ``Environmental
effects in the formation and evolution of extrasolar planetary
system''.  V.~N.~and G.~P.~acknowledge partial support by the
Universit\`a di Padova through the ``progetto di Ateneo \#CPDA103591''.
V.~G.~acknowledges support from PRIN INAF 2010 ``Planetary system at
young ages and the interactions with their active host stars''.
Some tasks of our data analysis have been carried out with
the VARTOOLS \citep{hartman2008} and \texttt{Astrometry.net} codes
\citep{lang2010}. This research has made use of the International 
Variable Star Index (VSX) database, operated at AAVSO, Cambridge, 
Massachusetts, USA.
\end{acknowledgements}

\bibliographystyle{aa}
\bibliography{biblio}

\begin{appendix}

\section{A new eclipsing variable}

A star in the WASP-3 field (UCAC3 $\alpha =
18^\mathrm{h}\,34^\mathrm{m}\,07^\mathrm{s}.36 $, $\delta =
+35^\circ\,38'\,59''.6 $, epoch 2000.0; \citealt{zacharias2010}),
initially chosen as reference star, was found to be variable and
excluded  from the reference list.  A complete light curve of this
variable was assembled by  registering the magnitudes on the
\texttt{N1-2} and \texttt{N4-N6} frames on a common zero point
(Fig.~\ref{var}, upper panel).  When folded on the $P\simeq0.3524$
days peak detected in the Lomb \& Scargle periodogram, the binned
curve shows a periodical pattern with equal maxima and slightly
different minima, typical of contact eclipsing binaries (W UMa-type;
Fig.~\ref{var}, lower panel). We derived the following ephemeris,
setting $\Phi=0$ at the phase of primary minimum and estimating
uncertainties through a bootstrapping algorithm:

\begin{eqnarray}
T_0 (\mathrm{BJD}_\mathrm{TDB})  &=& 2\,455\,776.48932\pm0.00024 + 
\nonumber \\ && + \,\Phi\cdot0.3523683\pm0.0000018
\end{eqnarray}


This variable star appears to be unpublished, and we submitted
it to the International Variable Star Index (identifier: VSX J183407.3+353859).
Colors from catalog
magnitudes:  $V=15.63$, $R=14.99$ (NOMAD; \citealt{zacharias2004}), 
$J=14.64$, $H=14.34$,
$K_\mathrm{s}=14.21$ (2MASS; \citealt{skru2006}), and proper motions
$\mu_\alpha\cos\delta=-32$ mas/yr, $\mu_\delta=19$ mas/yr (UCAC3)
suggest that this object could be a binary with both  components of
late-G spectral type. 

\begin{figure}[!t]
\centering \includegraphics[width=9cm]{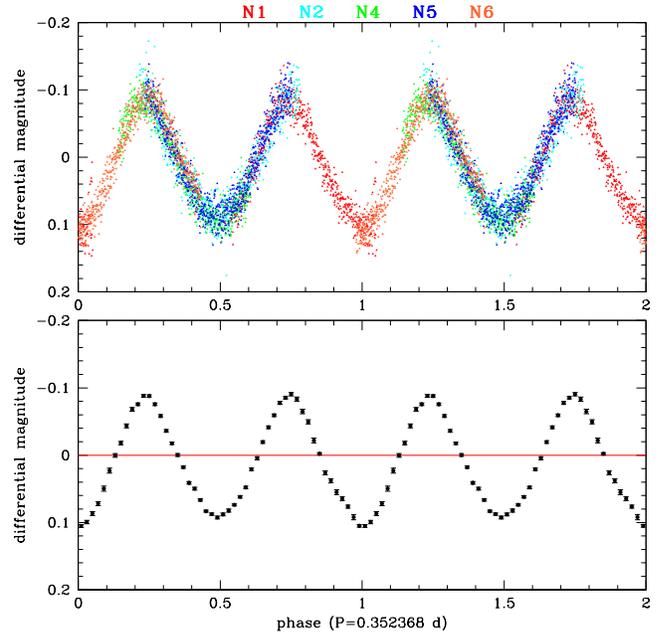}
\caption{Light curve of a previously unreported $R\sim15$ variable star
  in the WASP-3 field, classified  as a contact eclipsing binary (see
  text for details). \emph{Top panel:} unbinned data points folded  on
  the best-fit period $P=0.353626$ days. Different nights are coded in
  different colors. \emph{Bottom panel:} same as above, binned on 0.02
  intervals in phase.}
\label{var}
\end{figure}

\end{appendix}

\end{document}